\documentclass{appolb}
\usepackage{times}

\usepackage{epsf}
\def\beq{\begin{equation}}
\def\eeq{\end{equation}}
\def\bea{\begin{eqnarray}}
\def\eea{\end{eqnarray}}
\def\ve{\vert}
\def\vel{\left|}
\def\ver{\right|}
\def\nnb{\nonumber}
\def\ga{\left(}
\def\dr{\right)}

\def\rar{\rightarrow}
\def\nnb{\nonumber}
\def\la{\langle}
\def\ra{\rangle}
\def\ba{\begin{array}}
\def\ea{\end{array}}
\def\bea{\begin{eqnarray}}
\def\eea{\end{eqnarray}}

\begin{document}
\pagestyle{plain}
\newcount\eLiNe\eLiNe=\inputlineno\advance\eLiNe by -1
\title{Radiative dileptonic decays of B-meson  in the general two Higgs
doublet model%
\author{ G. Erkol and G. Turan
\address{Physics Department, Middle East Technical  University,
06531, Ankara, Turkey \\
E-mail: gurerk@newton.physics.metu.edu.tr \\
E-mail: gsevgur@metu.edu.tr }  }}
\maketitle
\begin{abstract}
We investigate the exclusive  $B\rightarrow \gamma \, \ell^+ \ell^- $ decay in
the general two  Higgs doublet model (model III) including the neutral Higgs
boson effects with an emphasis on possible CP-violating effects.
For this decay, we  analyse the dependencies of the forward-backward asymmetry of the lepton
pair, $A_{FB}$, CP-violating asymmetry, $A_{CP}$, and the CP-violating asymmetry 
in forward-backward asymmetry, $A_{CP}(A_{FB})$,  on the model 
parameters and also on the neutral Higgs  boson effects. We have found that 
$A_{FB}\sim 10^{-1}\, ,\, 10^{-2}$, $A_{CP}\sim 10^{-2}\, ,\, 10^{-1}$ and 
$A_{CP}(A_{FB}) \sim 10^{-2}\, ,\, 10^{-1}$ depending on the relative
magnitude of the Yukawa couplings $\bar{\xi}_{N,tt}^{U}$  and
$\bar{\xi}_{N,bb}^{D}$ in the model III. We also observe  that these
physical quantities   
are sensitive  to the model parameters and neutral Higgs boson effects are
quite sizable for some values of the coupling $\bar{\xi}_{N,\tau \tau}^{D}$. 

\noindent PACS number(s): 12.60.Fr, 13.20.He
\end{abstract} 
\section{Introduction}
It has been realized for a long time that rare B-meson decays  induced by the 
flavor--changing neutral currents (FCNC) are the most promising field for obtaining 
information about the fundamental parameters of the standard model (SM), like the 
elements of the Cabibbo--Kobayashi--Maskawa (CKM) matrix, the leptonic decay constants 
etc., and  testing the SM predictions at loop level. At the same time rare decays can also 
serve as a good probe for establishing new physics beyond the SM, such as the two Higgs 
doublet model (2HDM), minimal supersymmetric extension of the SM (MSSM) \cite{Hewett}, etc., 
since the contributions from these new models and the SM arise at the same order in 
perturbation theory. The observation of radiative penguin mediated processes, in both the 
exclusive $B \rar K^* \gamma$ \cite{Ammar} and inclusive $B \rar X_s \gamma$
\cite{CLEO} channels, stimulated 
the investigation of the radiative rare B meson decays with a new momentum. Among 
these rare decays,  $B \rar \gamma \, \ell^+ \ell^-$   $(\ell =e,\mu ,\tau)$
have received a 
special interest due to their relative cleanliness and sensitivity to new physics.
They have been investigated in the framework of the SM for light and heavy lepton 
modes in refs.\cite{Eilam1}-\cite{Aliev2}. The new physics effects in these decays have 
also been studied in some models, like MSSM \cite{Xiong} and the two Higgs doublet model
\cite{Iltan1,Aliev3,Erkol}. 

In this work, we study the radiative $B \rar \gamma \, \ell^+ \ell^- $ decay in
the general  two-Higgs doublet model (model III) including the neutral Higgs effects.
The 2HDM is the minimal extension of the SM, which consists of adding a
second doublet to the Higgs sector. In this model, the Yukawa Lagrangian responsible for
the interaction of quarks and leptons with gauge bosons opens up the
possibility of having tree-level FCNC, which are forbidden in the SM and
model I and II types of the 2HDM. This brings new parameters, i.e., Yukawa
couplings, into the theory.

$B \rar \gamma \,\ell^+ \ell^-$ decay is induced by the pure-leptonic decay  
$B \rar \ell^+ \ell^-$, which is free from
the helicity suppression, in contrast to the channels with light leptons, 
but  quite hard to detect experimentally due to low efficiency. In 
$B \rar \gamma\, \ell^+ \ell^-$ decay, 
helicity suppression is overcome by the photon emission in addition to the
lepton pair. For this reason, it is expected for $B \rar \gamma\, \ell^+ \ell^- $ 
decay  to have 
a large branching ratio and this makes its investigation  interesting. Another 
reason that motivates to study $B \rar \gamma \,\ell^+ \ell^-$  process is that it 
receives additional contributions from the neutral Higgs boson (NHB) exchanges in
the 2HDM. Since NHB contributions are proportional to either the lepton
mass or the corresponding Yukawa coupling, they are negligible for  $B \rar
\gamma \, \ell^+ \ell^-$ decays with light leptons, but  
we could expect  significant contributions for $\ell =\tau$. Indeed, 
the investigation of  $B \rar \gamma \, \tau^+ \tau^-$ decay in 
model I and II types of the 2HDM in \cite{Iltan1}, and in MSSM in \cite{Xiong}, including 
NHB effects report that  the contribution from exchanging neutral Higgs bosons may be quite
sizable for  large values of  $\tan \beta $, which is already favored by LEP experiments 
\cite{LEP}.

We investigated the $B \rar \gamma  \, \ell^+ \ell^-$ decay for $\ell=\tau$ in the 
model III type of the 2HDM in a previous paper \cite{Erkol}. In this work we extend this
study with an emphasis on possible CP violation effects. The CP asymmetry
is of great interest in high energy physics especially since its origin is still unclear.
In the SM, the source of CP violation is the complex CKM matrix elements,
which can explain all the existing data on CP violation. However, for
example,  to explain 
the matter-antimatter asymmetry observed in the universe today one needs additional sources 
of CP violating effects, which has motivated to search new models beyond the SM. 
In model III type of the 2HDM, the complex Yukawa couplings provide a possible source of 
CP violation. Indeed, it was reported \cite{Iltan2}-\cite{Iltan4} that a measurable  
CP aysmmetry was obtained due to this new phase in the model III. 

The paper is organized as follows: In section \ref{s2}, we first present the
leading order (LO) QCD corrected effective Hamiltonian for the quark level
process $b \rar  \gamma \,\ell^+ \ell^- $, including the NHB exchanges. Then we give 
the corresponding matrix element for the exclusive $B \rar \gamma \,\ell^+ \ell^-$
decay. Next, we calculate the  the forward-backward asymmetry of the lepton
pair, $A_{FB}$, CP-violating asymmetry, $A_{CP}$, and the CP-violating
asymmetry in forward-backward asymmetry, $A_{CP}(A_{FB})$, as functions of
the model parameters. Section \ref{s3} is devoted to the numerical analysis  
of these physical quantities with respect to the CP parameter $\sin \theta$, Yukawa
couplings $\bar{\xi}_{N,\tau \tau}^{D}$  and $\bar{\xi}_{N,bb}^{D}$ and the
mass ratio $m_{h^0}/m_{A^0}$ and to the discussion of our results. In
Appendix A, we give a brief summary about the general 2HDM (model III). 
The operators and  the corresponding  Wilson coeffients appearing in the effective 
Hamiltonian are given in Appendices B and C, respectively. Finally, some parametrizations 
used  in the text may  be found in  Appendix D.

\section{The exclusive  $B \rar \gamma \,\ell^+ \ell^-$ decay \label{s2}}
The exclusive $B \rar \gamma \,\ell^+ \ell^-$ decay is induced by the inclusive 
$b \rar  s \,\gamma \,\ell^+ \ell^- $  one. Therefore, we start with the QCD 
corrected effective Hamiltonian for the related quark level process
$b \rar s \,  \ell^+ \ell^-$, which is obtained
by integrating out heavy particles in the SM and in the 2HDM
\cite{Dai,Kruger}
\beq
{\cal H}=\frac{-4\, G_F}{ \sqrt{2}} V_{tb} V_{ts}^*
\Bigg{\{}\sum_{ i =1}^{10} C_{i}(\mu ) O_{i}(\mu)+
\sum_{ i =1}^{10} C_{Q_{i}}(\mu )Q_{i}(\mu) \Bigg{\}} \, ,
\label{H1}
\eeq
where $O_{i}$ are current-current $(i=1,2)$, penguin $(i=1,..,6)$, 
magnetic penguin $(i=7,8)$ and semileptonic $(i=9,10)$ operators . The 
additional operators $Q_{i}, (i=1,..,10)$ are due to the 
NHB exchange diagrams, which give considerable contributions in the case that the 
lepton pair is $\tau^{+}\tau^{-}$ \cite{Dai}. 
$C_{i}(\mu )$ and $C_{Q_{i}}(\mu )$ are Wilson coefficients renormalized at 
the scale $\mu$. All these operators and the Wilson coefficients, together 
with their initial values calculated at $\mu=m_W$ in the SM and also the 
additional coefficients coming from the new Higgs scalars are presented in 
Appendices B and C.

Neglecting the strange quark mass, the effective Hamiltonian (\ref{H1}) leads to 
the following matrix element for  $b \rar s \,  \ell^+ \ell^-$, 
\bea
{\cal M} & = & \frac{\alpha G_F}{ \sqrt{2}\, \pi} V_{tb} V_{ts}^*
 \Bigg{\{} C_9^{eff} (\bar s \gamma_\mu P_L b) \, \bar \ell \gamma_\mu \ell +
C_{10} ( \bar s \gamma_\mu P_L b) \, \bar \ell \gamma_\mu \gamma_5 \ell    \nnb \\
& -& 2 C_7 \frac{m_b}{q^2} (\bar s i \sigma_{\mu \nu} q_\nu P_R b) \bar \ell \gamma_\mu \ell 
 + C_{Q_{1}}(\bar s  P_R b) \bar \ell  \ell +C_{Q_{2}}(\bar s P_R b) \bar
 \ell \gamma_5 \ell \Bigg{\}}~,
\eea
where $P_{L,R}=(1\mp \gamma_5)/2$ , $q$ is the momentum transfer and $V_{ij}$'s are the 
corresponding elements of the CKM matrix.  

In order to obtain the matrix element for $b \rightarrow s \, \gamma\, \ell^{+}\ell^{-}$ decay, 
a photon line should be attached to any charged internal or external line. As pointed out 
before \cite{Aliev1,Aliev2},
contributions coming from the release of the free photon from  any charged internal line will 
be suppressed by a factor of $m^2_b/M^2_W$ and we  neglect them in the following analysis. When 
a photon is released from   the initial quark lines it contributes to the so-called 
"structure dependent" (SD) part of the amplitude. Using the expressions
\bea
\label{mel1}
\la \gamma(k) \vel \bar s \gamma_\mu
(1 \mp \gamma_5) b \ver B(p_B) \ra &=&
\frac{e}{m_B^2} \Big\{
\epsilon_{\mu\nu\lambda\sigma} \varepsilon^{\ast\nu} q^\lambda
k^\sigma g(q^2) \pm i\,
\Big[ \varepsilon^{\ast\mu} (k q) -
(\varepsilon^\ast q) k^\mu \Big] f(q^2) \Big\}~, \nnb 
\eea
\bea                                             
\la \gamma \ve \bar s i \sigma_{\mu \nu} k_\nu (1\mp \gamma_5) b \ve
B \ra &=& \frac{e}{m_B^2} \Bigg{\{} \epsilon_{\mu \alpha \beta \sigma}
\epsilon_\alpha^*
k_\beta q_\sigma \, g_1(p^2)
\mp ~ i \left[ \epsilon_\mu^* (k q) - (\epsilon^* k ) q_\mu \right] \,
f_1(p^2) \Bigg{\}}~,\label{ff2}\nnb \\ 
\la \gamma \ve \bar s (1+ \gamma_5 ) b \ve B \ra &=& 0~, 
\eea
the SD part of the amplitude  can be written as
\bea
\label{Msd}
{\cal M}_{SD} &=& \frac{\alpha G_F}{2 \sqrt{2} \, \pi} V_{tb} V_{ts}^* 
\frac{e}{m_B^2} \,\Bigg\{
\bar \ell \gamma^\mu  \ell \, \Big[
A_1 \epsilon_{\mu \nu \alpha \beta} 
\varepsilon^{\ast\nu} q^\alpha k^\beta + 
i \, A_2 \Big( \varepsilon_\mu^\ast (k q) - 
(\varepsilon^\ast q ) k_\mu \Big) \Big] \nnb \\
&+& \bar \ell \gamma^\mu \gamma_5 \ell \, \Big[
B_1 \epsilon_{\mu \nu \alpha \beta} 
\varepsilon^{\ast\nu} q^\alpha k^\beta 
+ i \, B_2 \Big( \varepsilon_\mu^\ast (k q) - 
(\varepsilon^\ast q ) k_\mu \Big) \Big] \Bigg\} \, ,
\eea
where $\varepsilon_\mu^\ast$ and $k_\mu$ 
are the four  vector polarization and four momentum of the photon, respectively, 
and $p_B$ is the momentum of the $B$ meson. In Eq. (\ref{Msd}),
$A_1$, $A_2$, $B_1$ and $B_2$ are functions of the Wilson coefficients and the
form factors, and they are given in Appendix D.

We note that the neutral Higgs exchange interactions do not contribute to the structure 
dependent part of the  matrix element ${\cal M}_{SD}$. However, the situation is different 
for the so-called "internal  Bremsstrahlung" (IB) contribution,
${\cal M}_{IB}$, which arises when a photon is radiated from one of  the final
$\ell$- leptons. Using the expressions
\bea
\la 0 \ve \bar s \gamma_\mu \gamma_5 b \ve B \ra &=& -~i f_B P_{B\mu}~, \nnb \\
\la 0 \ve \bar s \sigma_{\mu\nu} (1\pm\gamma_5) b \ve B \ra &=& 0~, \nnb \\
\la 0 \ve \bar s \gamma_5 b \ve B \ra &=& i f_B \frac{m_{B}^{2}}{m_{b}}~, \nnb \\
\la 0 \ve \bar s  b \ve B \ra &=& 0~,
\eea
and the conservation of the vector current, IB part of the matrix element is found as 
\cite{Iltan1}
\bea
\label{Mib}
{\cal M}_{IB} &=& \frac{\alpha G_F}{2 \sqrt{2} \, \pi} V_{tb} V_{ts}^*  
e f_B i \,\Bigg\{
F\, \bar \ell  \Bigg(
\frac{{\not\!\varepsilon}^\ast {\not\!p}_B}{2 p_1 k} - 
\frac{{\not\!p}_B {\not\!\varepsilon}^\ast}{2 p_2 k} \Bigg) 
\gamma_5 \ell \nnb \\
&+& F_1 \, \bar \ell \Bigg[
\frac{{\not\!\varepsilon}^\ast {\not\!p}_B}{2 p_1 k} -
\frac{{\not\!p}_B {\not\!\varepsilon}^\ast}{2 p_2 k} +
2 m_\ell \Bigg(\frac{1}{2 p_1 k} + \frac{1}{2 p_2 k}\Bigg)
{\not\!\varepsilon}^\ast \Bigg] \ell \Bigg\}~,
\eea
where $F$ and $F_1$ are functions of form factors and the Wilson coefficients $C_{Q_1}$ and
$C_{Q_2}$ due to the NHB effects and their explicit forms can be found in Appendix D.
Finally, the total matrix element for the $B \rar \gamma \, \ell^{+}\ell^{-}$ decay
is obtained as a sum of the ${\cal M}_{SD}$ and ${\cal M}_{IB}$ terms,
${\cal M}={\cal M}_{SD}+{\cal M}_{IB}$.

Now, we will calculate the forward-backward asymmetry, $A_{FB}$, for the lepton pair, 
CP-violating asymmetry, $A_{CP}$, and CP violating asymmetry in the forward-backward 
asymmetry, $A_{CP}(A_{FB})$  for the process under consideration. All these measurable
physical quantities can provide a great deal of clues to test  the theoretical models used.
We first give the definitions of $A_{FB}(x)$ and $A_{CP}$:
\begin{eqnarray} 
A_{FB}(x)& = & \frac{ \int^{1}_{0}dz \frac{d^2 \Gamma }{dx dz} - 
\int^{0}_{-1}dz \frac{d^2 \Gamma }{dx dz}}{\int^{1}_{0}dz 
\frac{d^2 \Gamma }{dx dz}+ \int^{0}_{-1}dz \frac{d^2 \Gamma }{dx dz}}~~, \label{AFB1} \\
A_{CP}& = & \frac{\Gamma (B\rar \gamma \, \ell^+ \ell^-) -\Gamma (\bar{B}\rar
\gamma \, \ell^+ \ell^-) }{\Gamma (B\rar \gamma \, \ell^+ \ell^-)  +
\Gamma (\bar{B}\rar \gamma \, \ell^+ \ell^-)} ~~, \label{ACP1}
\end{eqnarray}
where $z=\cos \theta $,  $\theta$ is the angle between the
momentum of the B-meson and that of $\ell^-$ and $x=2 E_{\gamma}/m_B$ is the
dimensionless photon energy.  In Eq.(\ref{AFB1}),   $\frac{d^2 \Gamma }{dx dz}$ 
is the double differential decay rate and in the center of mass (CM) frame of the 
dileptons $\ell^+\ell^-$, it is given by
\bea
\frac{d^2 \Gamma}{dx dz} & = & \vel \frac{\alpha G_F}{2 \sqrt{2} \, \pi} V_{tb}
V_{ts}^*  \ver^2 \,
\frac{\alpha}{\ga 2 \, \pi \dr^3}\,\frac{\pi}{4}\,m_B \, x \, v \, \nnb \\
& & \Bigg\{ \frac{m^2_B}{32} \, x^2 \,
\Bigg[((1+z^2)(1-x-4 r)) ( \vel A_1 \ver^2 + \vel A_2 \ver^2 +\vel B_1 \ver^2 + 
\vel B_2 \ver^2 ) \nnb \\ 
& + & 8 r (\vel A_1 \ver^2 + \vel A_2 \ver^2 )
+ 4 z \sqrt{(1-x)(1-x-4 r)}\mbox{\rm Re}
(A_2 B^*_1+A_1 B^*_2)\Bigg ] \nnb \\ 
& + & f_B m_{\ell} \frac{(x-1)}{((z^2-1)(x-1)+4 r z^2)}\Bigg[ v x z 
\mbox{\rm Re} (B_2 F^*-B_1 F^*_1) \nnb \\ & + & (1-4 r-z^2 (1-x-4 r))
\mbox{\rm Re}(A_2 F^*_1)-
x \mbox{\rm Re}(A_1 F^*)\Bigg ] \nnb \\ & + & 
f^2_B \frac{(1-x)}{x^2 ((z^2-1)(x-1)+4 r z^2)^2}
\Bigg [ \vel F \ver^2  \Big((-2+4 x-3 x^2+x^3 \nnb \\ & + & 
8 r (1-x)) (z^2-1)+4 r x^2 z^2  \Big) 
+ \vel F_1 \ver^2 \Big(\Big(32 r^2 (x-1)+4 r (4-6 x+2 x^2) \nnb \\
& - & 2+4 x-3 x^2+x^3
\Big)(z^2-1)+x^2 z^2 \Big) \Bigg ] \Bigg\}\, . 
\eea
where  $v=\sqrt{1-\frac{4 r}{1-x}}$ with  $r=m^2_{\ell}/m^2_B$.
Integrating over the angle variable, we find the forward backward asymmetry $A_{FB}$ as 
follows,
\begin{eqnarray}
A_{FB} & = & -\int\, dx\, 4 \, v \, x^2 \Bigg{\{} m^2_B \, x \, \sqrt{(x-1) (x-1+4 r)}
\mbox{\rm Re} (A_1 A^*_2-B_1 B^*_2) \nnb \\
& -& 4 f_B m_{\ell} \, v \, \Bigg(\frac{x-1}{x-1+4 r}\Bigg) \ln \frac{4 r}{1-x}
\mbox{\rm Re}((A_2- B_2) F^*\nnb \\ & - &  
(A_1- B_1) F^*_1) \Bigg{\}} \Bigg{/}\int\, dx\,D(x)\, , \label{AFB2}
\end{eqnarray}
where
\bea
D(x) & = & \frac{m_B^2 }{12} x^3 v\, \Bigg[ (\vel A_1 \ver^2 + \vel A_2 \ver^2
)(1+2 r-x)+
(\vel B_1 \ver^2 + \vel B_2 \ver^2 )(1-4 r-x)\Bigg] \nnb \\
& -& f_B m_{\ell} \, x \, \Bigg[ 2 v (1-x) \mbox{\rm Re}(A_2 F^*_1)+{\rm ln}
\frac{1 + v}{1-v}\Bigg( (x-4 r)\mbox{\rm Re} (A_2 F^*_1)-x \mbox{\rm Re}(A_1 F^*)  \Bigg )
\Bigg ]\nnb \\
& - & 2 f^2_B \Bigg[v \frac{(1-x)}{x} \Bigg( \vel F \ver^2+
(1-4 r) \vel F_1 \ver^2 \Bigg)
+ \, {\rm ln}\frac{1 + v}{1 - v} \Bigg(  \Bigg( 1 +\frac{2 r}{x} -
\frac{1}{x} -
\frac{x}{2} \Bigg)\vel F \ver^2 \nnb \\
& + & \Bigg(  (1-4 r) - \frac{2 \ga 1- 6 r + 8 r^2 \dr}{x}
 -\frac{x}{2}\Bigg) \vel F_1 \ver^2 \Bigg)\Bigg]  ~.\label{Dx}
\eea
We note that in these integrals the Dalitz boundary for the dimensionless photon energy 
$x$ is taken as 
\bea
\delta \leq x \leq 1-\frac{4 m_\ell^2}{ m^2_B}~, \label{KR}
\eea
since $\vel {\cal M}_{IB} \ver^2 $ term has infrared singularity due to
the emission of soft photon. In order to obtain a finite result, we follow the approach
described in ref.\cite{Aliev2} and impose a cut on the photon energy, i.e., we require 
$E_{\gamma}\geq 50$ MeV, which corresponds to detecting only hard photons experimentally. 
This cut requires that $E_{\gamma}\geq \delta \, m_B /2$ with $\delta =0.01$.

For  $B\rightarrow \gamma \, \ell^+ \ell^- $ decay, $A_{CP}$ almost
vanishes in the SM due to the unitarity of CKM matrix together with the
smallness of $V_{ub} V^*_{us}$. However, in model III complex Yukawa
couplings provide a  new source of CP violation. In our calculations, we
choose $\bar{\xi}^D_{N,bb}=|\bar{\xi}^D_{N,bb}| e^{i\theta}$ so that
$C^{eff}_{9}$, $C_{Q_1}$ and $C_{Q_2}$ are the Wilson coefficients that
contain CP violating terms. Using  Eq.(\ref{ACP1}), we calculate $A_{CP}$ as 
\begin{eqnarray}
A_{CP} & = & \frac{\int\, dx \, T(x)}{\int\, dx\,(D(x)+D_{CP}(x))}\, ,
\end{eqnarray}
where
\begin{eqnarray}
T(x)& = & m^2_B~~ x^2 ~~ \mbox{\rm Im}(\bar{\xi}^D_{N,bb}) ~~\Bigg{\{} \frac{v}{3}
x (1+2 r-x) A^{(2)}_1 A^{(3)}_1 \nnb \\ 
& &  \! \! \! \! \! \! \! \! \! \! \! \! \! \! \! \!\! \! \! \! \! \! \! \!
- 2 f_B \frac{m_{\ell}}{m_b}\Bigg[
\left( 2 v (1-x)+(x-4 r) {\rm ln} \frac{1 + v}{1-v} \right)
  A^{(2)}_2 F^{(2)}_1  - x {\rm ln} \frac{1 + v}{1-v} A^{(2)}_1 F^{(2)} \Bigg] 
\Bigg{\}} \, ,\nnb \\ & &  \label{Tx}
\end{eqnarray}
and $D_{CP}(x)$ is the CP conjugate of $D(x)$ which is defined  as 
\begin{eqnarray}
D_{CP}(x) & = & D(x)\left( \bar{\xi}^D_{N,bb}\rightarrow
(\bar{\xi}^D_{N,bb})^* \right) \, .
\end{eqnarray}
The explicit form of the functions $A^{(2)}_{(1)}$, $A^{(3)}_{(1)}$, etc.,
in Eq.(\ref{Tx}) are given in Appendix D.

Finally, we consider the CP violating asymmetry in $A_{FB}$, $A_{CP}
(A_{FB})$, which is an important measurable quantity that may provide
information about the model used. It  is defined as 
\begin{eqnarray}
A_{CP}(A_{FB})& = & \frac{A_{FB} -\bar{A}_{FB}}{A_{FB} +\bar{A}_{FB}} ~~, \label{ACPAFB1}
\end{eqnarray}
Here, $A_{FB}$ is given by Eq.(\ref{AFB2}) and $\bar{A}_{FB}$ is obtained by the replacement
$\bar{\xi}^D_{N,bb}\rightarrow (\bar{\xi}^D_{N,bb})^*$ in $A_{FB}$ .
%========================================================================
\section{Numerical analysis and discussion \label{s3}}
We present here our numerical results only for $\ell=\tau$ channel, but they
can easily be applied to the $\ell=\mu$ case.
The input parameters we used in our numerical analysis are as follows:
\begin{eqnarray}
& & m_B =5.28 \, GeV \, , \, m_b =4.8 \, GeV \, , \,m_c =1.4 \, GeV \, , \,
m_{\tau} =1.78 \, GeV \, , \nnb \\ 
& & m_{H^0} =150 \, GeV \, , \,m_{h^0} =70 \, GeV
\, , \,m_{A^0} =80 \, GeV \, , \, m_{H^{\pm}} =400 \, GeV \, , \nnb \\  
& & |V_{tb} V^*_{ts}|=0.045 \, , \, \alpha^{-1}=129 \,  , \,G_F=1.17 \times 10^{-5}\, GeV^{-2}
\,  , \,\tau_B=1.64 \times 10^{-12} \, s \,  , \nnb \\
& & C^{eff}_{9}=4.229 \,\, , \, \, C_{10}=-4.659.
\end{eqnarray}
Here we note that the value of the Wilson coefficient $C^{eff}_9$ above corresponds
to only the short-distance contributions. $C^{eff}_9$ also receives long-distance
(LD)   contributions associated with the real $\bar{c}c$ intermediate states (See 
Appendix C for the details of LD contributions). There are five possible resonances in
the $\bar{c}c$ system that can contribute to the $B \rar \gamma \,\tau^{+}\tau^{-}$ decay 
and to calculate their contributions  we need to  divide the integration region for x 
into two parts:
$\delta \leq x \leq 1-((m_{\psi_2}+0.02)/m_B)^2$ and
$1-((m_{\psi_2}-0.02)/m_B)^2 \leq x \leq 1-(2 m_{\tau}/m_B)^2 $, where 
$m_{\psi_2}=3.686 $ GeV is the mass of the second resonance.

For the values of the form factors $g,~f,~g_1$ and $f_1$, we have used the results of 
ref. \cite{Eilam2} and \cite{Buchalla}, and represent their $q^2$ dependencies in terms of 
two parameters $F(0)$ and $m_F$ as
\begin{eqnarray}
F(q^2) & = & \frac{F(0)}{\left(1-\frac{q^2}{m^2_F}\right)^2}
\end{eqnarray}
where the values  $F(0)$ and $m_F$ for the $B\rightarrow \gamma$
are listed in Table 1.

There are many free parameters in the general 2HDM, such as masses of the
charged and neutral Higgs bosons and complex Yukawa couplings,
$\xi_{ij}^{U,D}$, where $i,j$ are quark flavor indices. There are also some
experimental results that one can use to restrict these new parameters. In
this context, the stronger restriction comes from the analysis of the $\Delta
\, F =2$ decays with $F=K,B_d,D$ mesons, the $\rho$ parameter and the 
$B\rar X_s \, \gamma$ decay.

The contributions to the Wilson coefficient $C_7$ from the neutral Higgs
bosons $h^0$ and $A^0$ are given by \cite{Alil2}
\begin{eqnarray}
C^{H}_7 (m_W) & = & (V_{tb} V^*_{ts})^{-1}
\sum_{i=d,s,b}\,\bar{\xi}_{N,bi}^{D}\, \bar{\xi}_{N,is}^{D}\, \frac{Q_i}{8
m_i m_b} \,, \label{CHx} 
\end{eqnarray}
where $H=h^0,A^0$, and  $m_i$ and $Q_i$ are the masses and charges of the down quarks,
respectively. Eq. (\ref{CHx}) shows that the neutral Higgs bosons can give 
a large contribution to the Wilson coefficient $C_7$ and this contradicts 
with the CLEO data  \cite{CLEO} 
\bea
\label{cleomes}
BR(B \rar X_s \, \gamma ) & = & (3.15 \pm 0.35 \pm 0.32 ) \times 10^{-4} \,
.
\eea
However, assuming that the couplings $\bar{\xi}_{N,is}^{D}$ with $i=d,s,b$
and $\bar{\xi}_{N,bd}^{D}$ are small enough to reach the conditions 
$\bar{\xi}_{N,bb}^{D}\bar{\xi}_{N,is}^{D} \ll 1 $ and
$\bar{\xi}_{N,bd}^{D}\bar{\xi}_{N,ds}^{D} \ll 1 $ \cite{Alil2},
together with the constraints  from $\Delta
\, F =2$ mixing \cite{Aliev4} and  the $\rho$ parameter \cite{Atwood}, we obtain the
conditions 
\begin{eqnarray}
\bar{\xi}_{N,tc} &\ll&\bar{\xi}_{N,tt}^U \,\, ,\nnb \\
\bar{\xi}_{N,ib}^{D} & , & \bar{\xi}_{N,ij}^{D} \sim \, 0 \,i,j=d,s \, ,
\end{eqnarray}  
so that we only take into account $\bar{\xi}_{N,tt}^U$ and
$\bar{\xi}_{N,bb}^D$. In our work,   we choose $\bar{\xi}_{N,tt}^U$ as real
and $\bar{\xi}_{N,bb}^D$ as complex, namely
$\bar{\xi}_{N,bb}^D=|\bar{\xi}_{N,bb}^D| e^{i\theta}$.
As for the $\bar{\xi}_{N,\tau \tau}^U$, since it controls the contributions
due to the NHB effects, we leave $\bar{\xi}_{N,\tau \tau}^U$ as a free parameter to 
investigate the size of NHB effect on the  measurable quantities of the 
$B\rar \gamma \, \tau^+ \tau^-$ decay. 

\begin{figure}[htb]
\vskip -3.0truein
\centering
\epsfxsize=6.8in
\leavevmode\epsffile{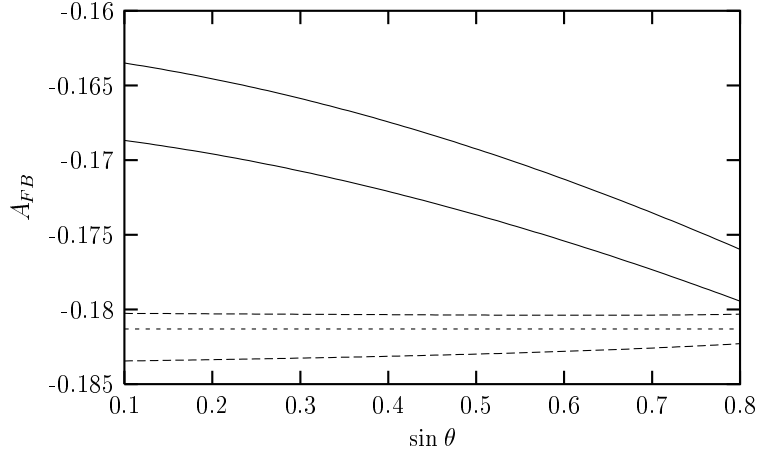}
\vskip -3.0truein
\caption[]{$A_{FB}$ as a function of $\sin \theta$ for $\bar{\xi}_{N,bb}^{D}=40\, m_b$
and $ |r_{tb}| <1$ without NHB effects. Here $A_{FB}$ is restricted in the
region between solid (dashed) curves for $C^{eff}_7 >0$ ($C^{eff}_7 <0$).
Dashed straight line represents the SM prediction. }
\label{AFBIII0sinrk1}
\end{figure}

\begin{figure}[htb]
\vskip -3.0truein
\centering
\epsfxsize=6.8in
\leavevmode\epsffile{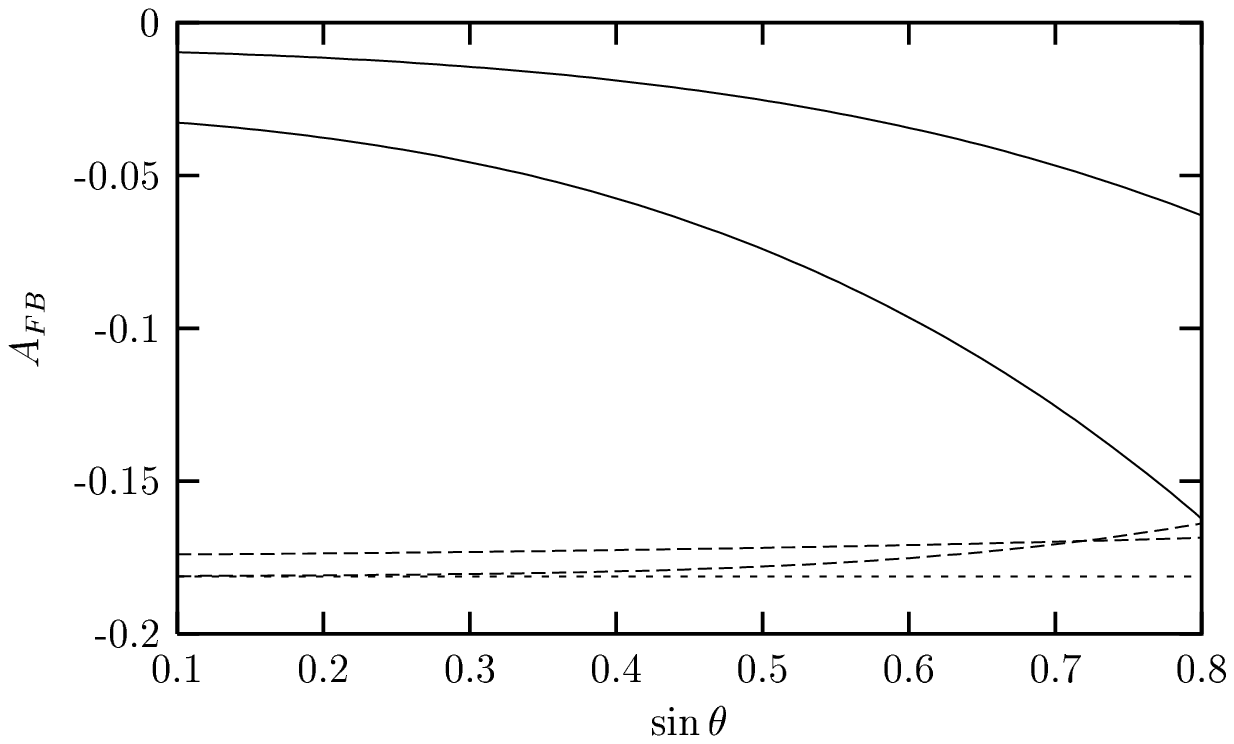}
\vskip -3.0truein
\caption[]{The same as Fig. \ref{AFBIII0sinrk1}, but including NHB effects
with $\bar{\xi}_{N,\tau\tau}^{D}= 10\, m_{\tau}$.}
\label{AFBIIINHBsinrk1}
\end{figure}

\begin{figure}[htb]
\vskip -3.0truein
\centering
\epsfxsize=6.8in
\leavevmode\epsffile{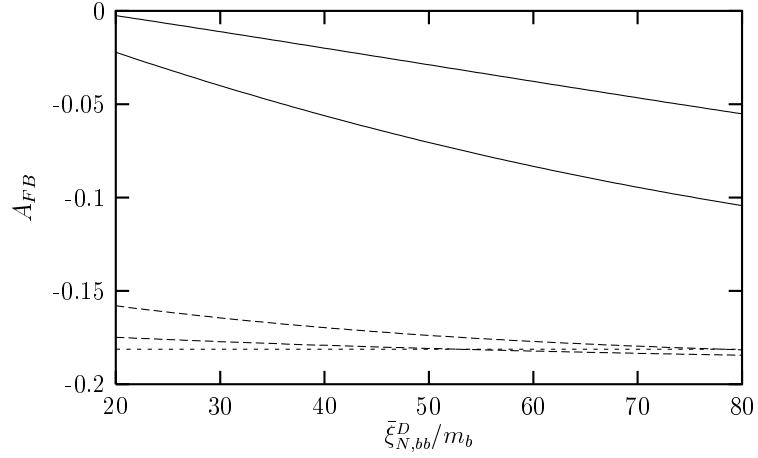}
\vskip -3.0truein
\caption[]{$A_{FB}$ as a function of  $\bar{\xi}_{N,bb}^{D}/m_{b}$  for
$\bar{\xi}_{N,\tau \tau}^{D}= 10 \, m_{\tau}$, $\sin \theta =0.5$ and $ |r_{tb}| <1$.
Here $A_{FB}$ is restricted in the
region between solid (dashed) curves for $C^{eff}_7 >0$ ($C^{eff}_7 <0$).
Dashed straight line represents the SM prediction. }
\label{AFBIIINHBkbbrk1}
\end{figure}

\begin{figure}[htb]
\vskip -3.0truein
\centering
\epsfxsize=6.8in
\leavevmode\epsffile{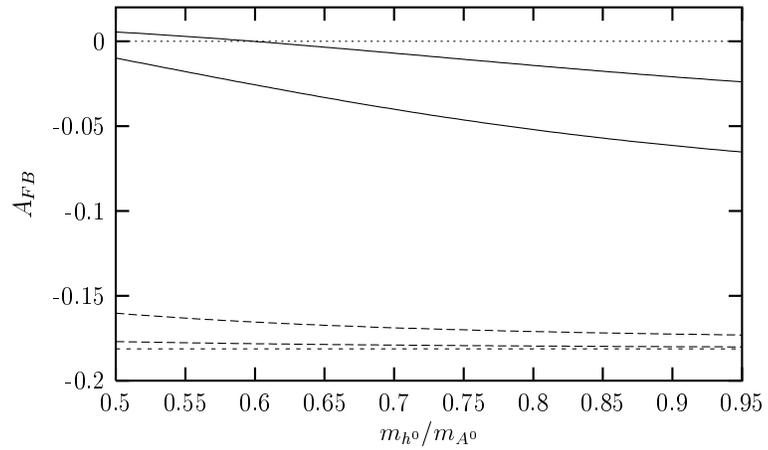}
\vskip -3.0truein
\caption[]{The same as Fig. \ref{AFBIIINHBkbbrk1}, but  as a function of
$m_{h^0}/m_{A^0}$.}  
\label{AFBIIINHBzhArk1}
\end{figure}

\begin{figure}[htb]
\vskip -3.0truein
\centering
\epsfxsize=6.8in
\leavevmode\epsffile{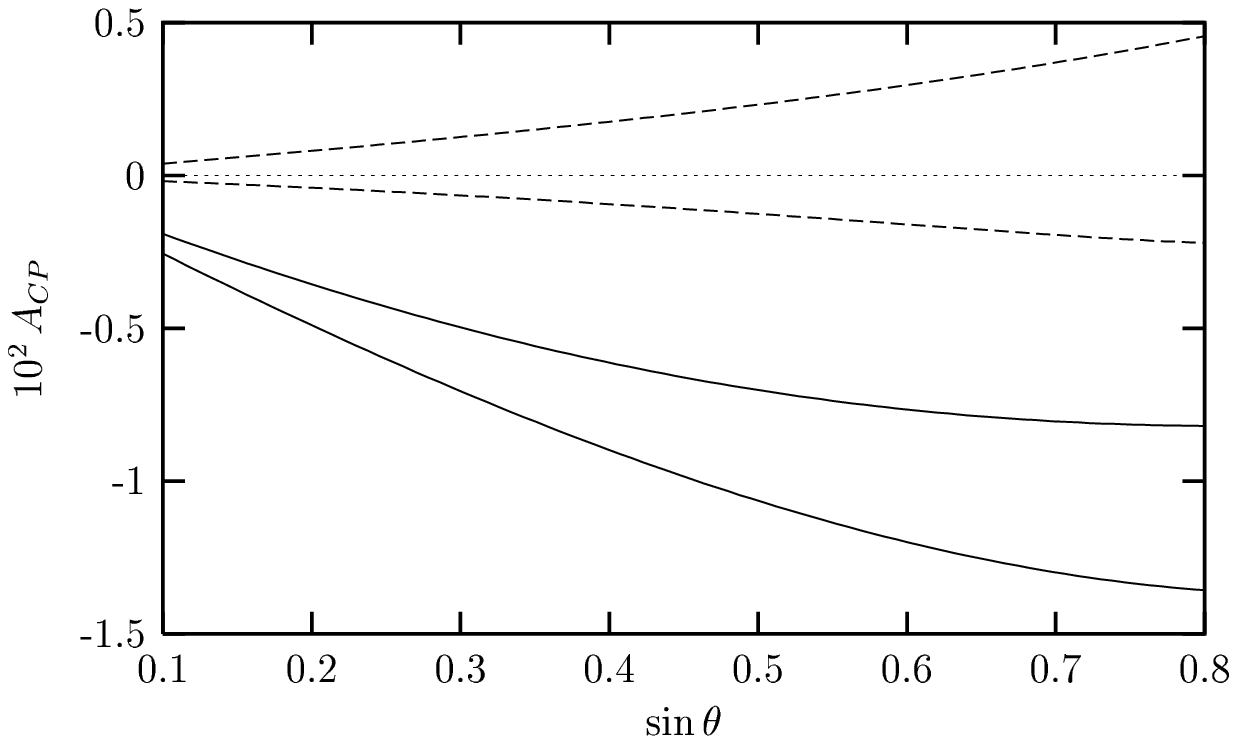}
\vskip -3.0truein
\caption[]{The same as Fig. \ref{AFBIII0sinrk1}, but for $A_{CP}$.} 
\label{ACPIIIsinrk1}
\end{figure}

\begin{figure}[htb]
\vskip -3.0truein
\centering
\epsfxsize=6.8in
\leavevmode\epsffile{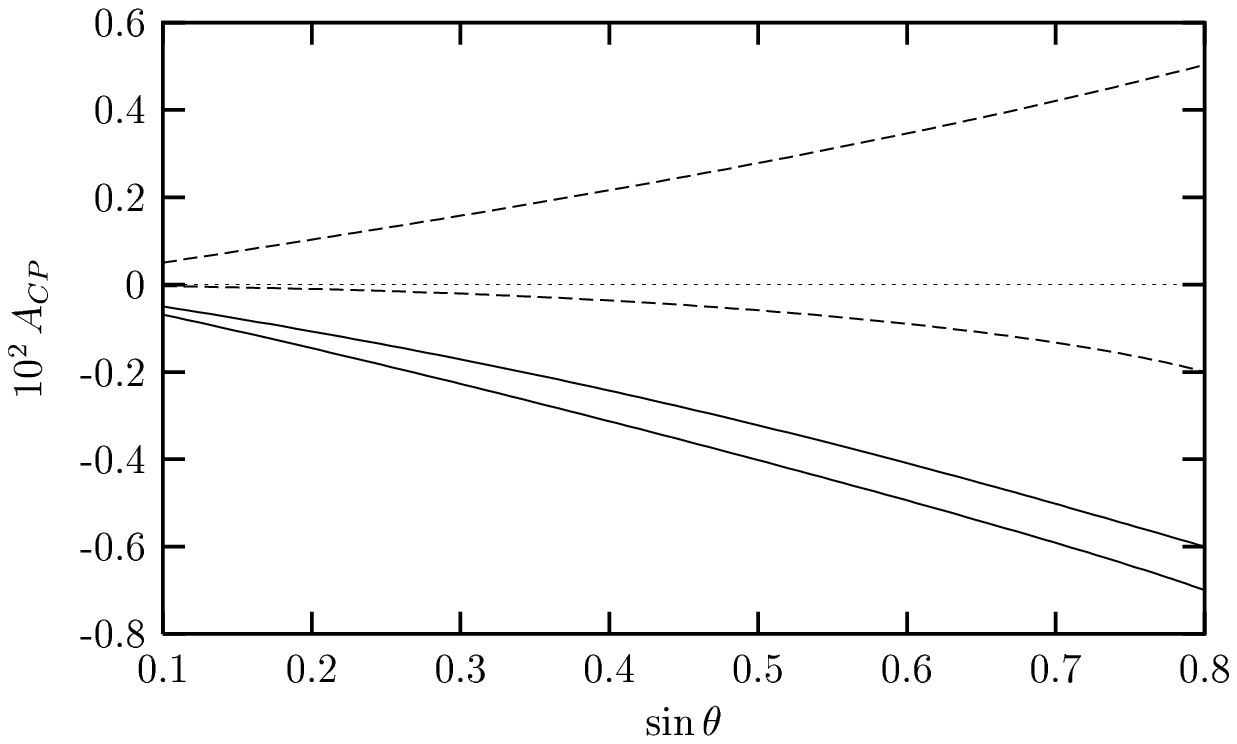}
\vskip -3.0truein
\caption[]{The same as Fig. \ref{AFBIIINHBsinrk1}, but for $A_{CP}$.}
\label{ACPIIINHBsinrk1}
\end{figure}

\begin{figure}[htb]
\vskip -3.0truein
\centering
\epsfxsize=6.8in
\leavevmode\epsffile{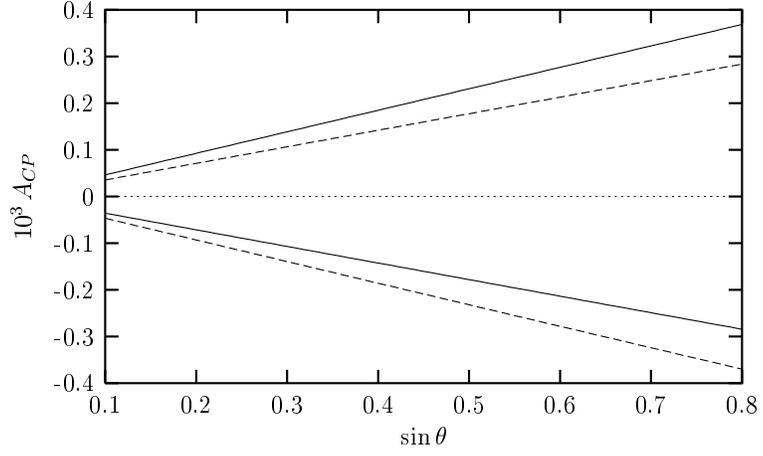}
\vskip -3.0truein
\caption[]{$A_{CP}$ as a function of   $\sin \theta$
for  $\bar{\xi}_{N,bb}^{D}=0.1\,m_b$ and  $r_{tb} >1$ without NHB effects.
Here $A_{CP}$ is restricted in the
region between solid (dashed) curves for $C^{eff}_7 >0$ ($C^{eff}_7 <0$).}
\label{ACPIIIsinrb1}
\end{figure}

\begin{figure}[htb]
\vskip -3.0truein
\centering
\epsfxsize=6.8in
\leavevmode\epsffile{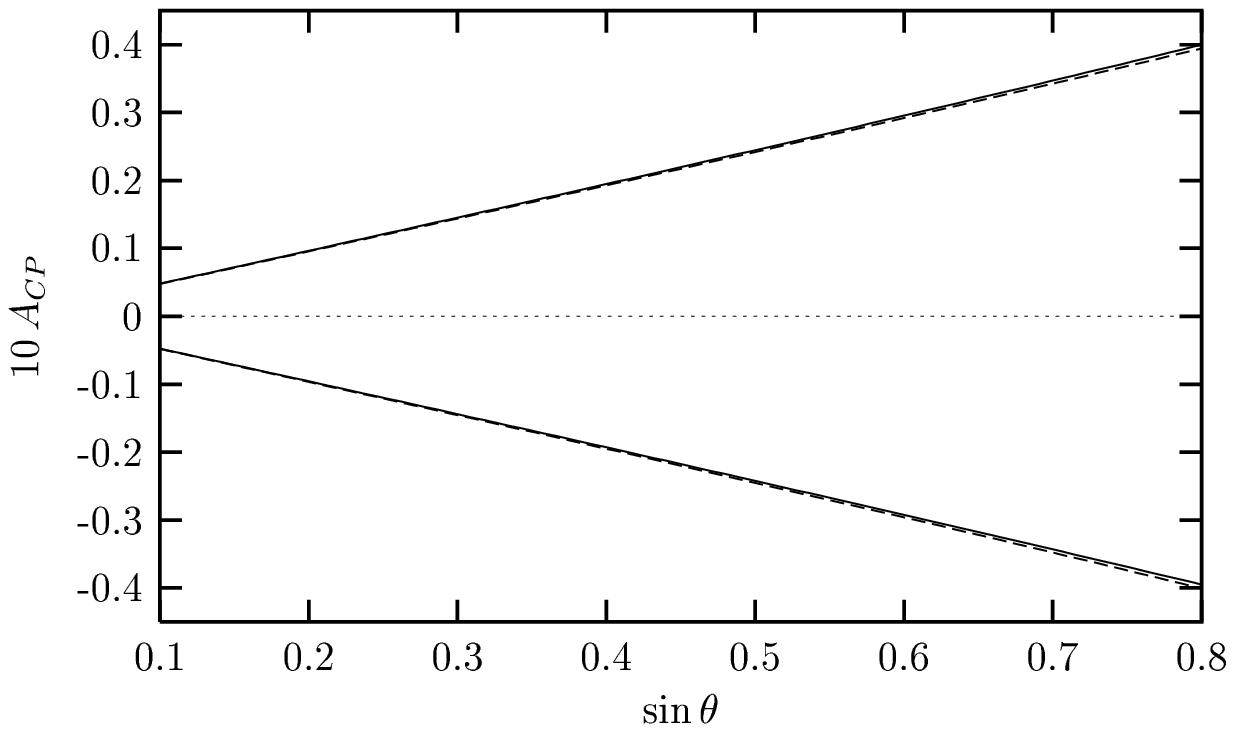}
\vskip -3.0truein
\caption[]{The same as Fig.\ref{ACPIIIsinrb1}, but including NHB effects
with $\bar{\xi}_{N,\tau\tau}^{D}=  m_{\tau}$.} 
\label{ACPIIINHBsinrb1}
\end{figure}

\begin{figure}[htb]
\vskip -3.0truein
\centering
\epsfxsize=6.8in
\leavevmode\epsffile{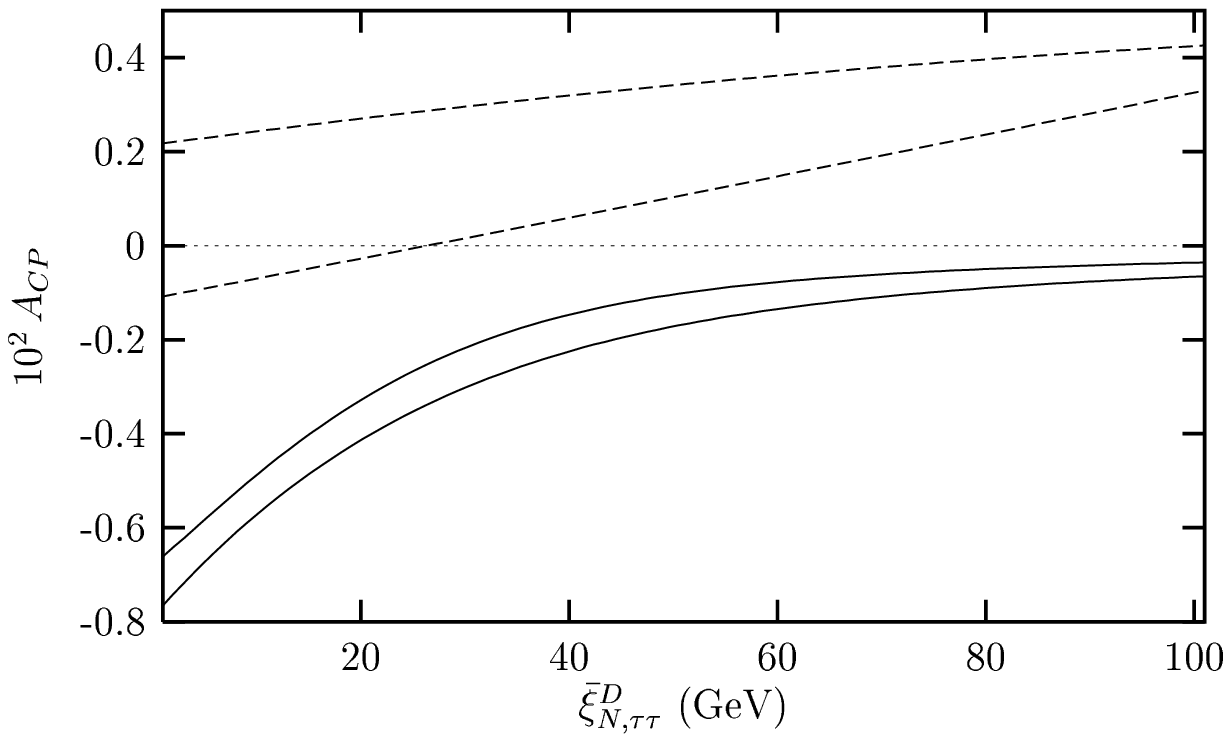}
\vskip -3.0truein
\caption[]{$A_{CP}$ as a function of  $\bar{\xi}_{N,\tau\tau}^{D}$  for
$\bar{\xi}_{N,bb}^{D}= 40 \, m_{b}$, $\sin \theta =0.5$ and $ |r_{tb}| <1$.
Here $A_{CP}$ is restricted in the
region between solid (dashed) curves for $C^{eff}_7 >0$ ($C^{eff}_7 <0$).}
\label{ACPIIINHBkttrk1}
\end{figure}

\begin{figure}[htb] 
\vskip -3.0truein   
\centering
\epsfxsize=6.8in    
\leavevmode\epsffile{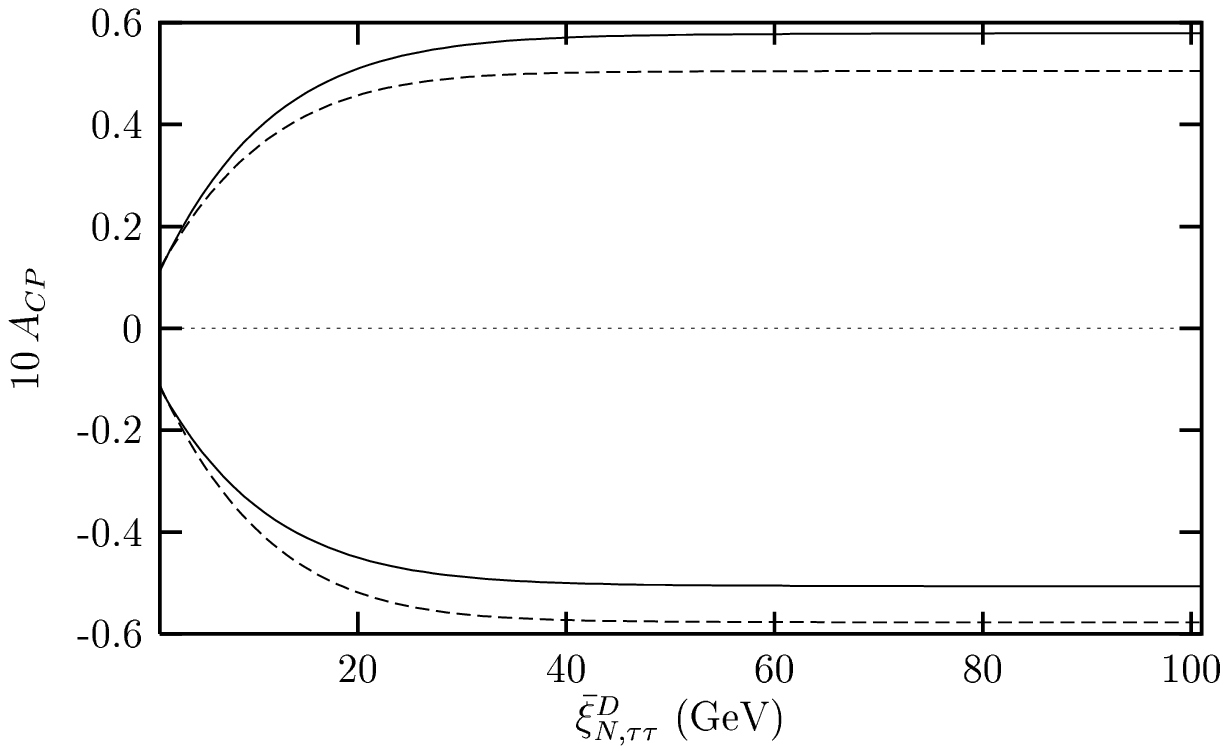}
\vskip -3.0truein   
\caption[]{The same as Fig. \ref{ACPIIINHBkttrk1},  but for $r_{tb}>1$ with 
$\bar{\xi}_{N,bb}^{D}= 0.1\, m_b$.}
\label{ACPIIINHBkttrb1}    
\end{figure}

\begin{figure}[htb]
\vskip -3.0truein
\centering
\epsfxsize=6.8in
\leavevmode\epsffile{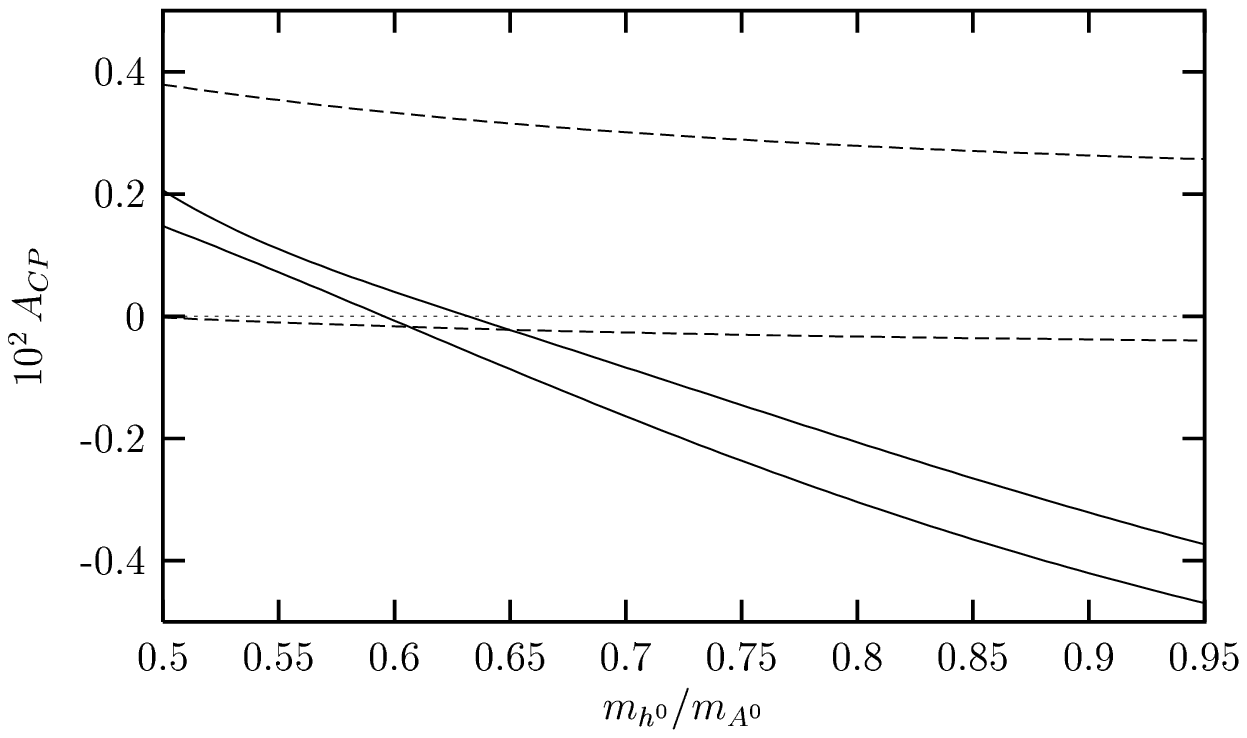}
\vskip -3.0truein
\caption[]{$A_{CP}$ as a function of  $m_{h^0}/m_{A^0}$  for
$\bar{\xi}_{N,\tau \tau}^{D}= 10 \, m_{\tau}$, $\bar{\xi}_{N,bb}^{D}= 40 \,
m_{b}$, $\sin \theta =0.5$ and $ |r_{tb}| <1$. Here $A_{CP}$ is restricted in the
region between solid (dashed) curves for $C^{eff}_7 >0$ ($C^{eff}_7 <0$).}
\label{ACPIIINHBzhArk1} 
\end{figure}

\begin{figure}[htb]
\vskip -3.0truein
\centering
\epsfxsize=6.8in  
\leavevmode\epsffile{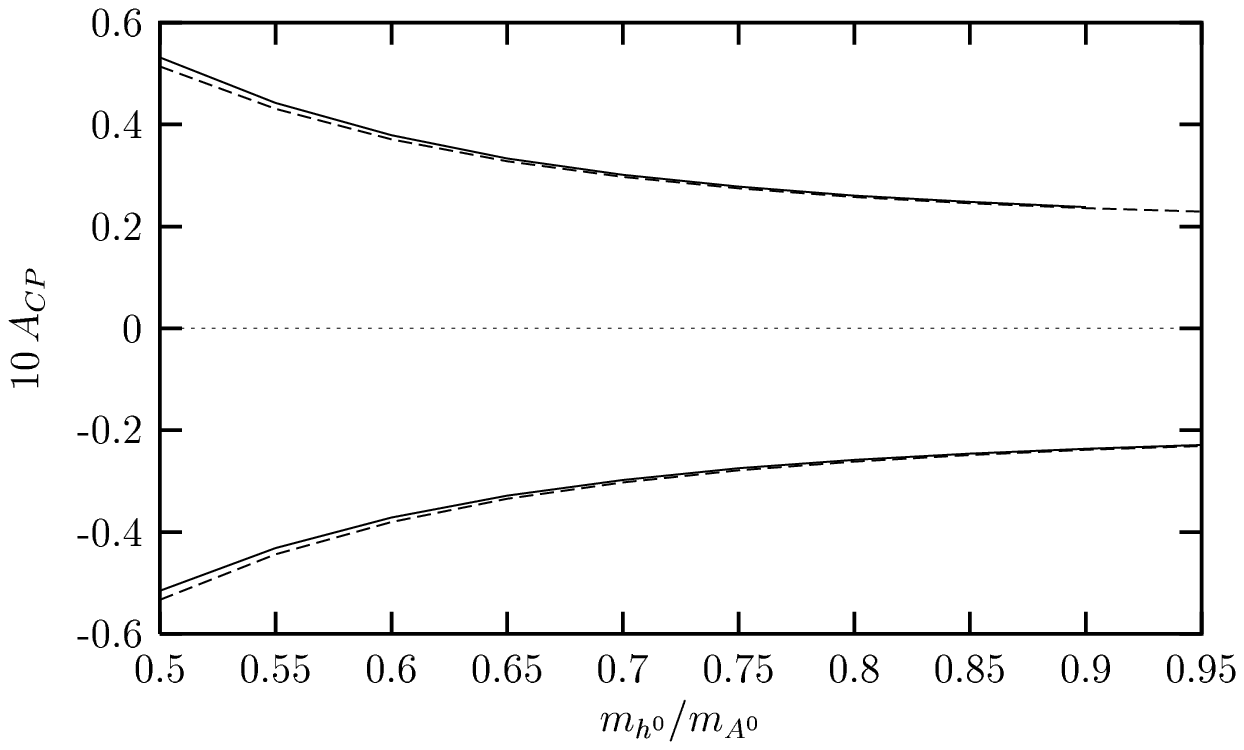}
\vskip -3.0truein
\caption[]{The same as Fig. \ref{ACPIIINHBzhArk1}, but for $r_{tb}>1$ with
$\bar{\xi}_{N,\tau \tau}^{D}=  m_{\tau}$ and $\bar{\xi}_{N,bb}^{D}= 0.1 \,
m_{b}$,.}
\label{ACPIIINHBzhArb1}
\end{figure}

\begin{figure}[htb]
\vskip -3.0truein
\centering
\epsfxsize=6.8in
\leavevmode\epsffile{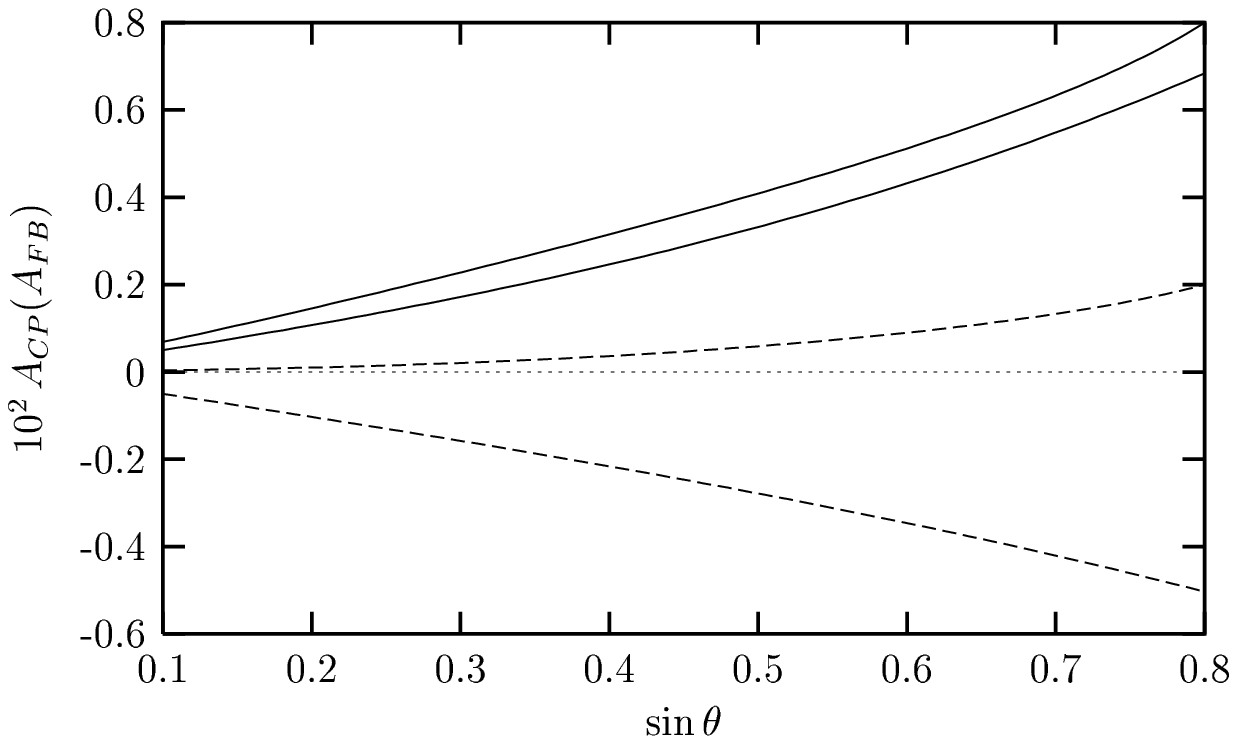}
\vskip -3.0truein
\caption[]{$A_{CP}(A_{FB})$ as a function of   $\sin \theta$
for $\bar{\xi}_{N,\tau \tau}^{D}= 10\, m_{\tau}$,  
$\bar{\xi}_{N,bb}^{D}=40\,m_b$ and  $|r_{tb}| <1$. Here $A_{CP}(A_{FB})$ is
restricted in the region between solid (dashed) curves for $C^{eff}_7 >0$ 
($C^{eff}_7 <0$).}
\label{AFBACPIIINHBsinrk1}
\end{figure}

\begin{figure}[htb]
\vskip -3.0truein
\centering
\epsfxsize=6.8in  
\leavevmode\epsffile{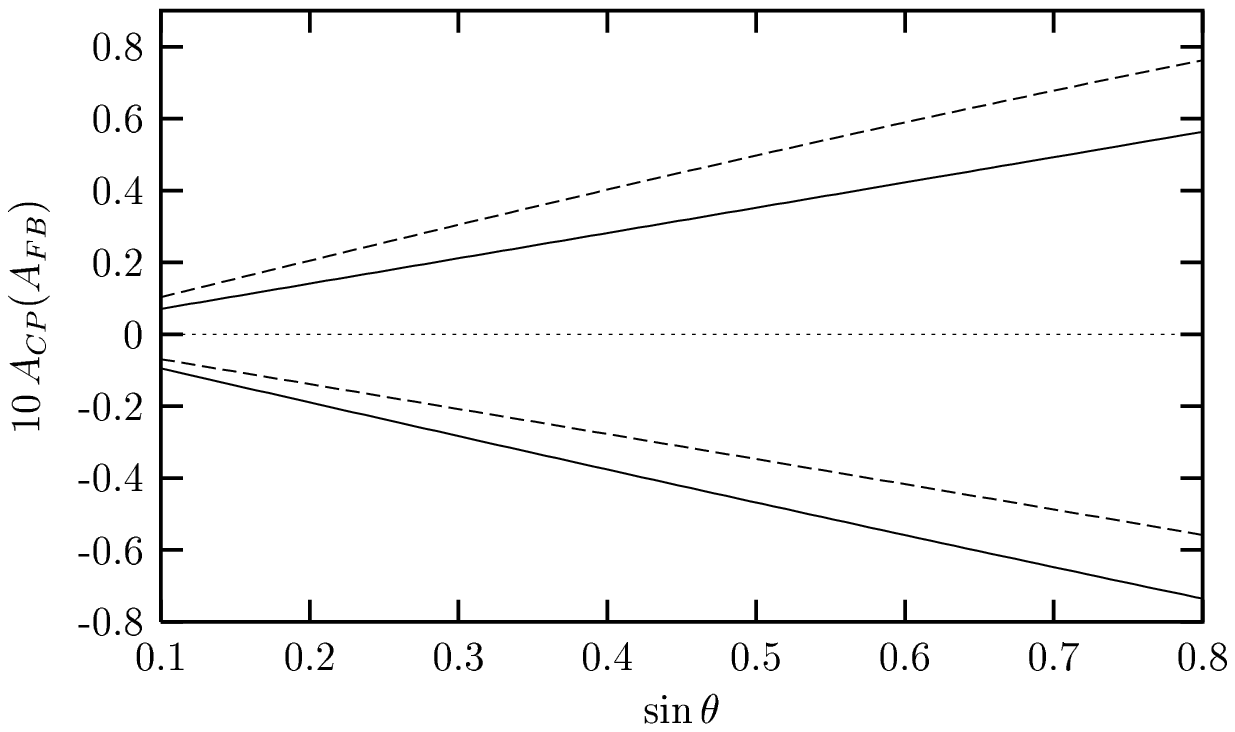}
\vskip -3.0truein
\caption[]{The same as Fig. \ref{AFBACPIIINHBsinrk1}, but for $r_{tb}>1$
with $\bar{\xi}_{N,\tau \tau}^{D}= m_{\tau}$ and $\bar{\xi}_{N,bb}^{D}=0.1\,m_b$. }
\label{AFBACPIIINHBsinrb1}
\end{figure}

\begin{figure}[htb]
\vskip -3.0truein
\centering       
\epsfxsize=6.8in
\leavevmode\epsffile{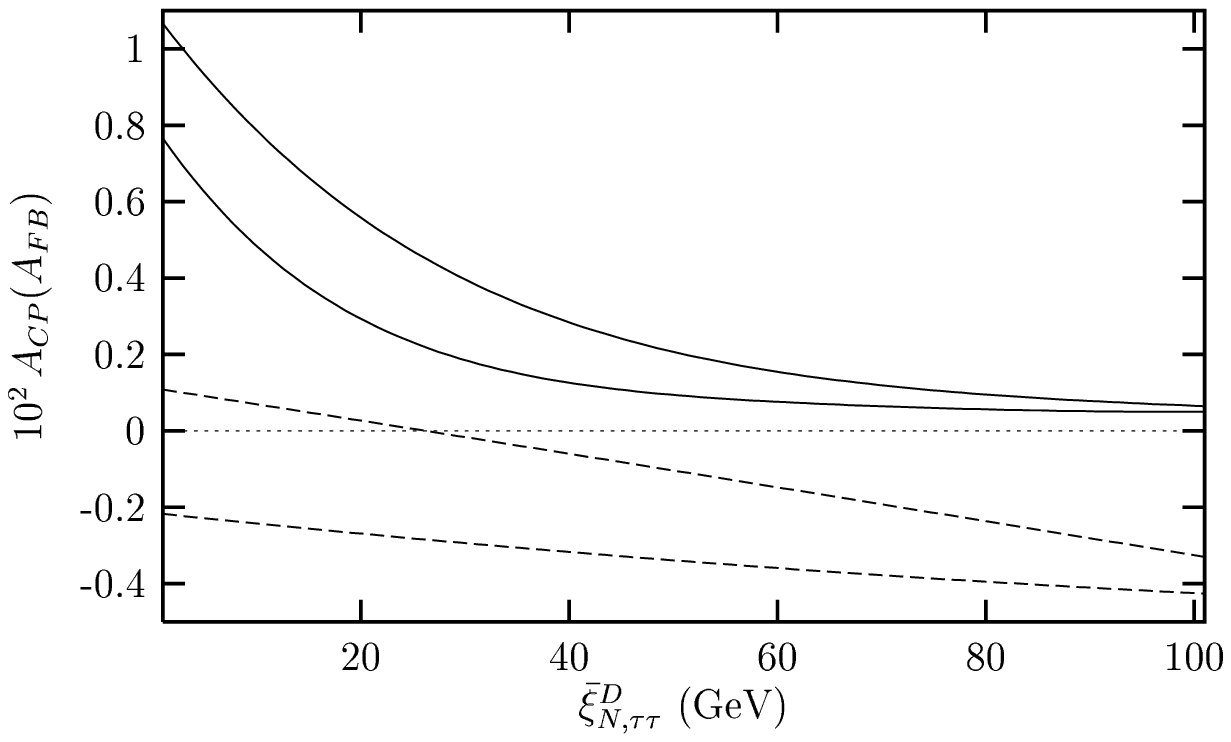}
\vskip -3.0truein
\caption[]{$A_{CP}(A_{FB})$ as a function of  $\bar{\xi}_{N,\tau\tau}^{D}$  for
$\bar{\xi}_{N,bb}^{D}= 40\, m_{b}$, $\sin \theta =0.5$
and $ |r_{tb}| <1$. Here $A_{CP}(A_{FB})$ is      
restricted in the region between solid (dashed) curves for $C^{eff}_7 >0$ 
($C^{eff}_7 <0$).}
\label{AFBACPIIINHBkttrk1}
\end{figure}

\begin{figure}[htb]
\vskip -3.0truein
\centering       
\epsfxsize=6.8in
\leavevmode\epsffile{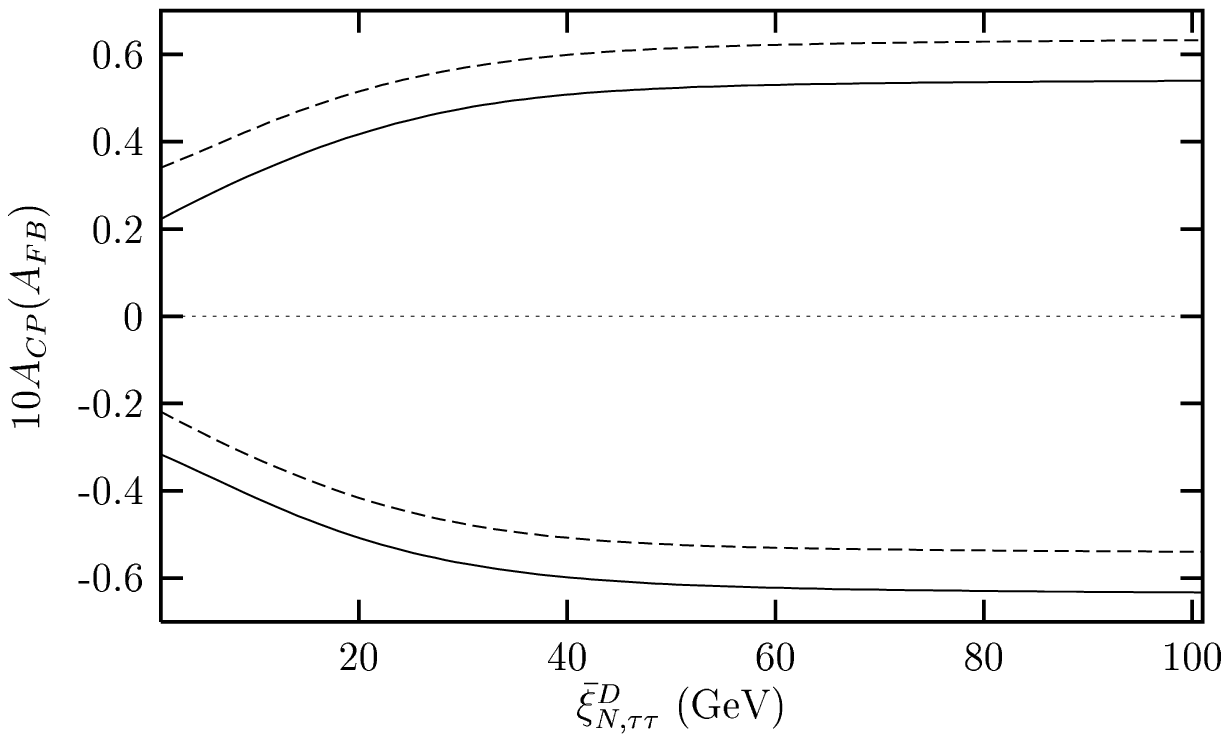}
\vskip -3.0truein
\caption[]{The same as Fig. \ref{AFBACPIIINHBkttrk1}, but for $r_{tb}>1$
with $\bar{\xi}_{N,bb}^{D}=0.1\,m_b$. }
\label{AFBACPIIINHBkttrb1}
\end{figure}

\begin{figure}[htb]
\vskip -3.0truein
\centering       
\epsfxsize=6.8in
\leavevmode\epsffile{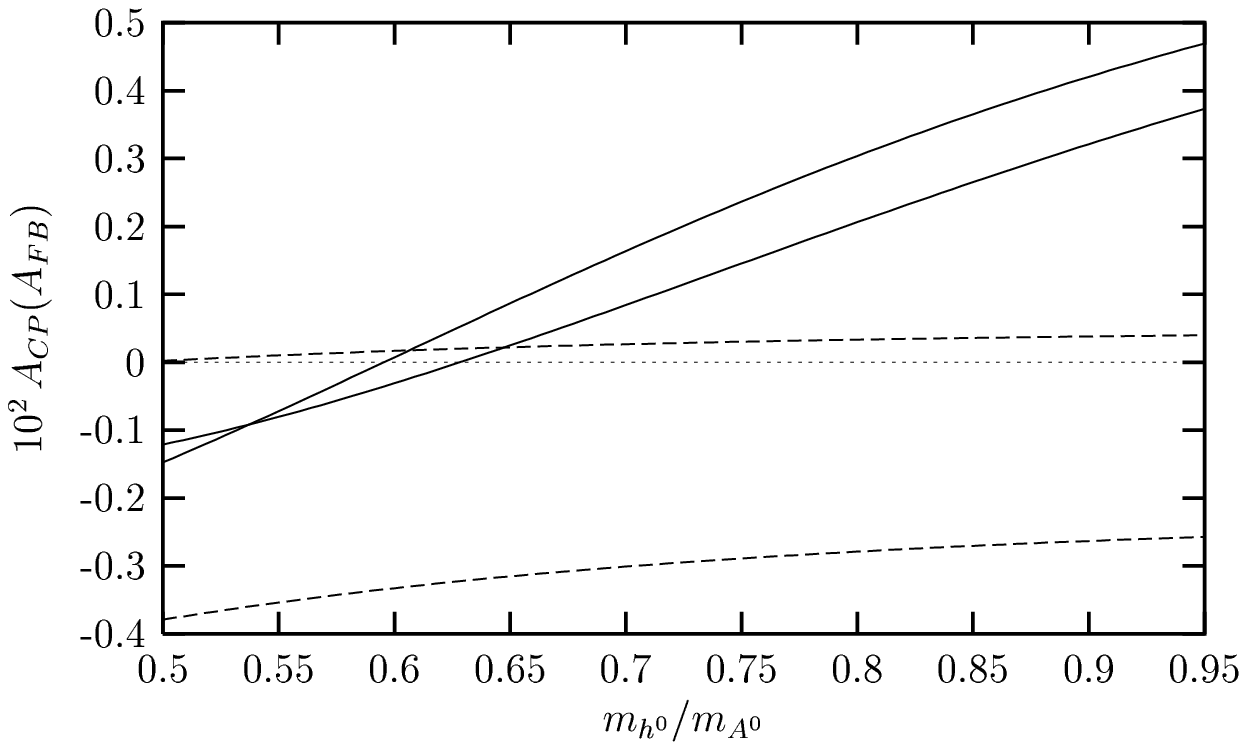}
\vskip -3.0truein
\caption[]{$A_{CP}(A_{FB})$ as a function of $m_{h^0}/m_{A^0}$ for
$\bar{\xi}_{N,\tau\tau}^{D}=10\, m_{\tau}$, $\bar{\xi}_{N,bb}^{D}= 40\,
m_{b}$, $\sin \theta =0.5$ and $ |r_{tb}| <1$. Here $A_{CP}(A_{FB})$ is
restricted in the region between solid (dashed) curves for $C^{eff}_7 >0$
($C^{eff}_7 <0$).}
\label{AFBACPIIINHBzhArk1}
\end{figure}

\begin{figure}[htb]
\vskip -3.0truein
\centering       
\epsfxsize=6.8in
\leavevmode\epsffile{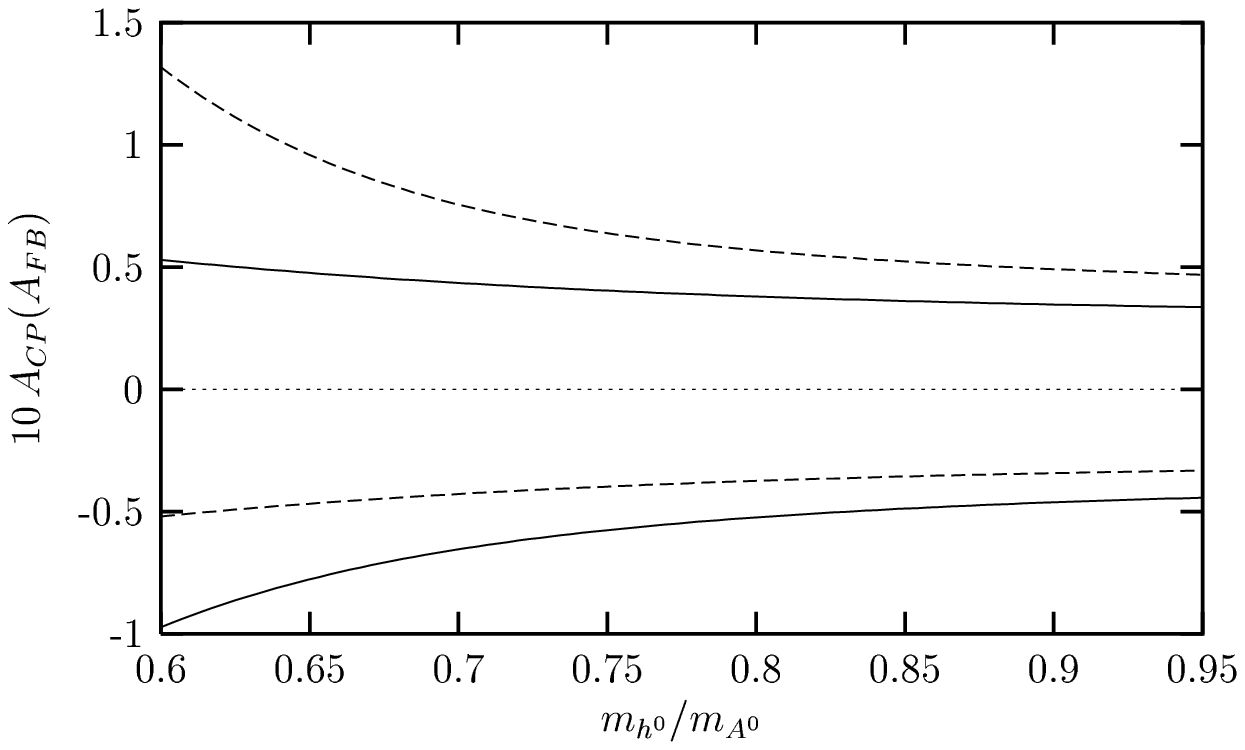}
\vskip -3.0truein
\caption[]{The same as Fig. \ref{AFBACPIIINHBzhArk1}, but for $r_{tb}>1$
with $\bar{\xi}_{N,\tau\tau}^{D}= m_{\tau}$ and  $\bar{\xi}_{N,bb}^{D}=0.1\,m_b$.}
\label{AFBACPIIINHBzhArb1}
\end{figure}

In our numerical calculations, we further  adopted the constraint on the Wilson
coefficient $C^{eff}_7$,  $0.257 \leq |C_7^{eff}| \leq 0.439$ \cite{Aliev4}
due to the CLEO measurement (Eq.(\ref{cleomes})) and  the redefinition
\begin{eqnarray}
\xi^{U,D}=\sqrt{\frac{4 G_F}{\sqrt{2}}} \bar{\xi}^{U,D} \nonumber \,\, .
\label{xineutr}
\end{eqnarray}
The above constraint on the $C^{eff}_7$ restricts the allowed regions of
the measurable quantities of the exclusive $B\rar \gamma \, \tau^+\tau^-$
decay, namely,
$A_{FB}$, $A_{CP}$ and $A_{CP}(A_{FB})$ and these regions are represented
by the ones between the solid curves for $C^{eff}_7>0$  and the dashed curves
for $C^{eff}_7<0$ throughout the graphs in Figs.
\ref{AFBIII0sinrk1}-\ref{AFBACPIIINHBzhArb1}.

In Fig. \ref{AFBIII0sinrk1}, we plot $\sin \theta$ dependence of $A_{FB}$ 
without NHB effects for the case of ratio $ |r_{tb}|\equiv \vel
\frac{\bar{\xi}_{N,tt}^U}{\bar{\xi}_{N,bb}^D}\ver <1$. 
We see that in model III without NHB effects,  $|A_{FB}|$  is smaller
compared to  its value in the SM (0.183), represented by the dashed straight
line, for  $C^{eff}_7>0$, but it
can be enhanced up to $7\%$ with increasing $\sin \theta$. For $C^{eff}_7<0$,  
$A_{FB}$ is not much sensitive to  $\sin \theta$, but its value  can be slightly
greater than the SM prediction. Including the NHB effects to $A_{FB}$ (
Fig. \ref{AFBIIINHBsinrk1}) reduces its magnitude $30\%$ 
of its value without NHB effects for $C^{eff}_7>0$, while for
$C^{eff}_7<0$, $A_{FB}$ is almost the same as the SM value.

In Fig. \ref{AFBIIINHBkbbrk1} (\ref{AFBIIINHBzhArk1}), we present $A_{FB}$ as a
function of $\bar{\xi}_{N,bb}^{D}$ ($m_{h^0}/m_{A^0}$) for
$\bar{\xi}_{N,\tau\tau}^{D}=10\, m_{\tau}$, $\sin \theta=0.5$ and
$ |r_{tb}| <1$. 
$A_{FB}$ is at the order of magnitude $10^{-1}$ and increases with the increasing 
values of both $\bar{\xi}_{N,bb}^{D}$ and $m_{h^0}/m_{A^0}$. For $C^{eff}_7>0$, 
$A_{FB}$ stands less than the SM prediction and for $C^{eff}_7<0$,
model III prediction can reach the SM one for large values
of  $\bar{\xi}_{N,bb}^{D}$ and $m_{h^0}/m_{A^0}$. 

We have also calculated the $\sin \theta$ ( $\bar{\xi}_{N,bb}^{D}$
and $m_{h^0}/m_{A^0}$) dependence of $A_{FB}$
for the case of ratio $r_{tb}>1$. We have found that in this case $A_{FB}$
is one (two) order(s) of magnitude smaller than its value for $ |r_{tb}| <1$ case,
and including NHB effects reduces this value even one more order of
magnitude. Therefore we do not present these graphs here.

Fig. \ref{ACPIIIsinrk1} represents $\sin \theta $ dependence of $A_{CP}$ without NHB
effects for the case of ratio $ |r_{tb}| <1$. 
It is at the order of magnitude $10^{-2}$ and increases
with $\sin \theta $. For $C^{eff}_7 <0$, $A_{CP}$ can have both signs, while
for $C^{eff}_7>0$ its sign does not change in the restricted region.
Including the NHB effects (Fig. \ref{ACPIIINHBsinrk1}) reduces $|A_{CP}|$
without NHB effects almost by $60\%$ for $C^{eff}_7>0$. However, for
$C^{eff}_7 <0$, it is possible to enhance it by up to $35\%$.

Figs. \ref{ACPIIINHBsinrb1} and \ref{ACPIIIsinrb1} are devoted to 
$\sin \theta $ dependence of $A_{CP}$ for $r_{tb}>1$ with and without NHB effects, 
respectively. Without NHB effects, $|A_{CP}|$  is at the order of magnitude
$10^{-3}$ and including the NHB effects can enhance it up to two orders of
magnitude, i.e., it becomes  $|A_{CP}|\sim 10^{-1}$. 
We further note that the restricted region for  $C^{eff}_7>0$ (solid lines )
and $C^{eff}_7 <0$  (dashed lines) are now larger but they almost
coincide.

Fig. \ref{ACPIIINHBkttrk1} (\ref{ACPIIINHBkttrb1}) shows $A_{CP}$ as a 
function of $\bar{\xi}_{N,\tau\tau}^{D}$ for  $\bar{\xi}_{N,bb}^{D}=40 (0.1) m_b$
and $\sin \theta=0.5$ for $ |r_{tb}| <1$ ($r_{tb}>1$). We see that $A_{CP}$
is sensitive to the parameter $\bar{\xi}_{N,\tau\tau}^{D}$ and it decreases
(increases) with the  increasing values of $\bar{\xi}_{N,\tau\tau}^{D}$ for
$C^{eff}_7>0$ ($C^{eff}_7 <0$) when $ |r_{tb}| <1$. When $r_{tb}>1$,
restricted regions for  $C^{eff}_7>0$ (solid lines ) and  $C^{eff}_7 <0$   
(dashed lines) almost coincide and $|A_{CP}|$ can take one orders of magnitude larger
values compared to the case where $ |r_{tb}| <1$.

In Fig. \ref{ACPIIINHBzhArk1} (\ref{ACPIIINHBzhArb1}), we plot the
dependence of $A_{CP}$ on $m_{h^0}/m_{A^0}$ for
$\bar{\xi}_{N,\tau\tau}^{D}=10 (1) \, m_{\tau}$,  $\bar{\xi}_{N,bb}^{D}=40 (0.1) m_b$
and $\sin \theta=0.5$ for $ |r_{tb}| <1$ ($r_{tb}>1$). For $ |r_{tb}| <1$
and  $C^{eff}_7>0$,
$A_{CP}$ is sensitive to the ratio $m_{h^0}/m_{A^0}$ and increases with the
increasing values of $m_{h^0}/m_{A^0}$. However, for $C^{eff}_7 <0$,
dependence of $A_{CP}$ on the ratio is weak but, the restricted region is
larger this time. As seen from Fig. \ref{ACPIIINHBzhArb1}, when $r_{tb}>1$,
 $|A_{CP}|$ can take one orders of magnitude larger values compared to the 
case where $ |r_{tb}|<1$. 

Finally, we present our results about the CP violating asymmetry in $A_{FB}$,
$A_{CP}(A_{FB})$ in a series of figures,
Figs.\ref{AFBACPIIINHBsinrk1}-\ref{AFBACPIIINHBzhArb1}.
Fig. \ref{AFBACPIIINHBsinrk1} (\ref{AFBACPIIINHBsinrb1}) shows $A_{CP}(A_{FB})$ as 
a function of $\sin \theta$ for $\bar{\xi}_{N,\tau\tau}^{D}=10 (1) \, m_{\tau}$,
$\bar{\xi}_{N,bb}^{D}=40 (0.1) m_b$ for $ |r_{tb}| <1$ ($r_{tb}>1$). For  $
|r_{tb}| <1$ ,  $A_{CP}(A_{FB})$ is at the 
order of magnitude $10^{-2}$ and it does not  change sign in the restricted
region for $C^{eff}_7>0$, while it can have both signs for $C^{eff}_7 <0$.  For  $ 
r_{tb}>1$ ,  $A_{CP}(A_{FB})$ can reach $5\%$ for the intermediate values of
$\sin \theta$ and restricted regions for
$C^{eff}_7>0$ (solid lines ) and  $C^{eff}_7 <0$ (dashed lines) almost coincide.
  
We can see from Figs.\ref{AFBACPIIINHBkttrk1} and \ref{AFBACPIIINHBkttrb1}
that $A_{CP}(A_{FB})$ is sensitive to the parameter
$\bar{\xi}_{N,\tau\tau}^{D}$, especially for its small values.  $A_{CP}(A_{FB})$ is
a decreasing (increasing) function of $\bar{\xi}_{N,\tau\tau}^{D}$ for 
$|r_{tb}| <1$ ($r_{tb}>1$) and reaches $1(6)\%$ for
$\bar{\xi}_{N,\tau\tau}^{D}=1 (50)$.

Fig. \ref{AFBACPIIINHBzhArk1} (\ref{AFBACPIIINHBzhArb1}) is devoted to
the ratio $m_{h^0}/m_{A^0}$ dependence of $A_{CP}(A_{FB})$ 
for $\bar{\xi}_{N,\tau\tau}^{D}=10 (1) \, m_{\tau}$,
$\bar{\xi}_{N,bb}^{D}=40 (0.1) m_b$ and $\sin \theta =0.5$ 
for $ |r_{tb}| <1$ ($r_{tb}>1$). It is seen from Fig.
\ref{AFBACPIIINHBzhArk1} that for $ |r_{tb}| <1$, $A_{CP}(A_{FB})$ is sensitive 
to the mass ratio $m_{h^0}/m_{A^0}$ and it is increasing with the increasing values of 
$m_{h^0}/m_{A^0}$ for $C^{eff}_7>0$, while for $C^{eff}_7 <0$, its
dependence on the mass ratio is weak. For $r_{tb}>1$, $|A_{CP}(A_{FB})|$ is
one order of magnitude larger than its value for  $ |r_{tb}| <1$.

In conclusion, we have investigated the physical quantities $A_{FB}$,
$A_{CP}$ and $A_{CP}(A_{FB})$ for the exlusive $B\rar \gamma \,
\ell^+\ell^-$ decay in the general 2HDM including the NHB effects. From the
results we have obtained we conclude that experimental investigation of
these quantities may be very  useful for testing the new physics effects
beyond the SM and also the sign of $C^{eff}_7$.

\begin{table}[h]                    
\renewcommand{\arraystretch}{1.5}                        
\addtolength{\arraycolsep}{3pt}
$$
\begin{array}{|l|cc|}
\hline
& F(0) & a_F \\ \hline
g &
\phantom{-}1 \, GeV & 5.6 \, GeV \\
f &
\phantom{-}0.8 \, GeV & 6.5 \, GeV \\
g_1 &
 \phantom{-}3.74 \, GeV^2 & 6.4 \, GeV \\
f_1&
  \phantom{-}0.68 \, GeV^2 & 5.5 \, GeV \\
\hline
\end{array}   
$$
\caption{$B$ meson decay form factors in the light-cone QCD sum rule.}
\renewcommand{\arraystretch}{1}
\addtolength{\arraycolsep}{-3pt}
\end{table}       
\newpage
\begin{appendix}
\section{Model Description}
The 2HDM is the minimal extension of the SM, which consists of adding a second 
doublet to the Higgs sector. In this model,  there are one charged Higgs scalar, 
two neutral Higgs scalars and one neutral Higgs pseudoscalar.
The general Yukawa Lagrangian, which is responsible for the
interactions of quarks  with gauge bosons,  can be written as
\begin{eqnarray}
{\cal{L}}_{Y}&=&\eta^{U}_{ij} \bar{Q}_{i L} \tilde{\phi_{1}} U_{j R}+
\eta^{D}_{ij} \bar{Q}_{i L} \phi_{1} D_{j R}+
\xi^{U\, \dagger}_{ij} \bar{Q}_{i L} \tilde{\phi_{2}} U_{j R}+
\xi^{D}_{ij} \bar{Q}_{i L} \phi_{2} D_{j R}
+ h.c. \,\,\, ,
\label{lagrangian}
\end{eqnarray}
where $i,j$  are family indices of quarks , $L$ and $R$ 
denote chiral projections $L(R)=1/2(1\mp \gamma_5)$, $\phi_{m}$ for $m=1,2$, 
are the two scalar doublets, $Q_{i L}$ are quark  
doublets, $U_{j R}$, $D_{j R}$ are the corresponding quark 
singlets, $\eta^{U,D}_{ij}$ and $\xi^{U,D}_{ij}$ are the matrices 
of the Yukawa couplings. 
The Yukawa Lagrangian in Eq. (\ref{lagrangian}) opens up the possibility 
that there appear tree-level FCNC, which are forbidden in the SM and 
model I and model II types of the 2HDM. However, tree-level FCNC
are permitted in the general 2HDM, and this type of 2HDM is referred to as model III 
in the literature.

In this model, it is possible to choose $\phi_1$ and $\phi_2$ in the following form
\begin{eqnarray}
\phi_{1}=\frac{1}{\sqrt{2}}\left[\left(\begin{array}{c c} 
0\\v+H^{0}\end{array}\right)\; + \left(\begin{array}{c c} 
\sqrt{2} \chi^{+}\\ i \chi^{0}\end{array}\right) \right]\, ; 
\phi_{2}=\frac{1}{\sqrt{2}}\left(\begin{array}{c c} 
\sqrt{2} H^{+}\\ H_1+i H_2 \end{array}\right) \,\, ,
\label{choice}
\end{eqnarray}
with the vacuum expectation values,  
\begin{eqnarray}
<\phi_{1}>=\frac{1}{\sqrt{2}}\left(\begin{array}{c c} 
0\\v\end{array}\right) \,  \, ; 
<\phi_{2}>=0 \,\, .
\label{choice2}
\end{eqnarray}
With this choice, the SM particles can be collected in the first doublet 
and the new particles in the second one. Further, we take $H_{1}$, $H_{2}$ 
as the mass eigenstates $h^{0}$, $A^{0}$ respectively. Note that, at tree 
level, there is no mixing among CP even neutral Higgs bosons, namely 
the SM one, $H^0$, and beyond, $h^{0}$.

The part which produces FCNC at tree level is  
\begin{eqnarray}
{\cal{L}}_{Y,FC}=
\xi^{U\,\dagger}_{ij} \bar{Q}_{i L} \tilde{\phi_{2}} U_{j R}+
\xi^{D}_{ij} \bar{Q}_{i L} \phi_{2} D_{j R} +
\xi^{D}_{kl} \bar{l}_{k L} \phi_{2} E_{l R} + h.c. \,\, .
\label{lagrangianFC}
\end{eqnarray}
In Eq.(\ref{lagrangianFC}), the couplings  $\xi^{U,D}$ for the
flavor-changing charged interactions are 
\begin{eqnarray}
\xi^{U}_{ch}&=& \xi_{neutral} \,\, V_{CKM} \nonumber \,\, ,\\
\xi^{D}_{ch}&=& V_{CKM} \,\, \xi_{neutral} \,\, ,
\label{ksi1} 
\end{eqnarray}
where  $\xi^{U,D}_{neutral}$ 
is defined by the expression
\begin{eqnarray}
\xi^{U (D)}_{N}=(V_{R(L)}^{U (D)})^{-1} \xi^{U,(D)} V_{L(R)}^{U (D)}\,, 
\label{ksineut}
\end{eqnarray}
and $\xi^{U,D}_{neutral}$ is denoted as $\xi^{U,D}_{N}$. 
Here the charged couplings are  the linear combinations of neutral 
couplings multiplied by $V_{CKM}$ matrix elements (see \cite{Aliev4} for
details). 
\section{The operator basis}
The operator basis in the general  2HDM (model III ) for our process  
is \cite{Dai,Kruger,Grinstein2,Misiak}
\begin{eqnarray}
& &  O_1 = (\bar{s}_{L \alpha} \gamma_\mu c_{L \beta})
               (\bar{c}_{L \beta} \gamma^\mu b_{L \alpha})\, \, , \, \,
     O_2 = (\bar{s}_{L \alpha} \gamma_\mu c_{L \alpha}) 
               (\bar{c}_{L \beta} \gamma^\mu b_{L \beta}) \, , \nnb \\  
& &  O_3 = (\bar{s}_{L \alpha} \gamma_\mu b_{L \alpha})
               \sum_{q=u,d,s,c,b}
               (\bar{q}_{L \beta} \gamma^\mu q_{L \beta}) \, \, ,  \, \,
     O_4 = (\bar{s}_{L \alpha} \gamma_\mu b_{L \beta})
                \sum_{q=u,d,s,c,b}
               (\bar{q}_{L \beta} \gamma^\mu q_{L \alpha}) \,\, , \nnb \\
& &  O_5 = (\bar{s}_{L \alpha} \gamma_\mu b_{L \alpha})
               \sum_{q=u,d,s,c,b}
               (\bar{q}_{R \beta} \gamma^\mu q_{R \beta}) \, \, , \, \,   
     O_6 = (\bar{s}_{L \alpha} \gamma_\mu b_{L \beta})
                \sum_{q=u,d,s,c,b}
               (\bar{q}_{R \beta} \gamma^\mu q_{R \alpha}),  \nnb   \\  
& &  O_7 = \frac{e}{16 \pi^2}
          \bar{s}_{\alpha} \sigma_{\mu \nu} (m_b R + m_s L) b_{\alpha}
                {\cal{F}}^{\mu \nu} \, \, , \,\,                           
     O_8 = \frac{g}{16 \pi^2}
                     \bar{s}_{\alpha} T_{\alpha \beta}^a \sigma_{\mu \nu} (m_b R +
                          m_s L) b_{\beta} {\cal{G}}^{a \mu \nu}  \,\, , \nnb \\
& & O_9 = \frac{e}{16 \pi^2} (\bar{s}_{L \alpha} \gamma_\mu b_{L \alpha})
              (\bar{\ell} \gamma^\mu \ell)  \,\, ,  \, \,
    O_{10} = \frac{e}{16 \pi^2}
          (\bar{s}_{L \alpha} \gamma_\mu b_{L \alpha})
              (\bar{\ell} \gamma^\mu \gamma_{5} \ell)  \,\, ,  \nnb \\
& & Q_1  =  \frac{e^2}{16
\pi^2}(\bar{s}^{\alpha}_{L}\,b^{\alpha}_{R})\,(\bar{\ell}\ell )
               \, \, , \, \, 
    Q_2 = \frac{e^2}{16 \pi^2}(\bar{s}^{\alpha}_{L}\,b^{\alpha}_{R})\,
               (\bar{\ell} \gamma_5 \ell ) \, \, , \label{op1} \\
& & Q_3 =\frac{g^2}{16 \pi^2}(\bar{s}^{\alpha}_{L}\,b^{\alpha}_{R})\,
          \sum_{q=u,d,s,c,b }(\bar{q}^{\beta}_{L} \, q^{\beta}_{R} ) \, \, , \, \, 
    Q_4 =\frac{g^2}{16 \pi^2}(\bar{s}^{\alpha}_{L}\,b^{\alpha}_{R})\,
             \sum_{q=u,d,s,c,b } (\bar{q}^{\beta}_{R} \, q^{\beta}_{L} )
               \,\, , \nnb \\
& & Q_5 =\frac{g^2}{16 \pi^2}(\bar{s}^{\alpha}_{L}\,b^{\beta}_{R})\,
          \sum_{q=u,d,s,c,b } (\bar{q}^{\beta}_{L} \, q^{\alpha}_{R} ) \, \, ,
               \, \,
    Q_6 =   \frac{g^2}{16 \pi^2}(\bar{s}^{\alpha}_{L}\,b^{\beta}_{R})\, ,
               \sum_{q=u,d,s,c,b } (\bar{q}^{\beta}_{R} \, q^{\alpha}_{L} ) \, , \nnb \\
& & Q_7 =   \frac{g^2}{16 \pi^2}(\bar{s}^{\alpha}_{L}\,\sigma^{\mu \nu} \, 
             b^{\alpha}_{R})\,
                  \sum_{q=u,d,s,c,b } (\bar{q}^{\beta}_{L} \, \sigma_{\mu \nu } 
                    q^{\beta}_{R} ) \, \, , \, \, 
      Q_8 =   \frac{g^2}{16 \pi^2}(\bar{s}^{\alpha}_{L}\,\sigma^{\mu \nu} 
                  \, b^{\alpha}_{R})\,
                  \sum_{q=u,d,s,c,b } (\bar{q}^{\beta}_{R} \, \sigma_{\mu \nu } 
                  q^{\beta}_{L} ) \, , \nnb \\ 
& & Q_9 =  \frac{g^2}{16 \pi^2}(\bar{s}^{\alpha}_{L}\,\sigma^{\mu \nu} 
               \, b^{\beta}_{R})\,
            \sum_{q=u,d,s,c,b }(\bar{q}^{\beta}_{L} \, \sigma_{\mu \nu } 
                    q^{\alpha}_{R} ) \, \, ,\, \,
        Q_{10}= \frac{g^2}{16 \pi^2}(\bar{s}^{\alpha}_{L}\,\sigma^{\mu \nu} \, 
b^{\beta}_{R})\,
\sum_{q=u,d,s,c,b }(\bar{q}^{\beta}_{R} \, \sigma_{\mu \nu } q^{\alpha}_{L})\nnb
\end{eqnarray}
where $\alpha$ and $\beta$ are $SU(3)$ colour indices and 
${\cal{F}}^{\mu \nu}$ and ${\cal{G}}^{\mu \nu}$ are the field strength 
tensors of the electromagnetic and strong interactions, respectively. Note 
that there are also flipped chirality partners of these operators, which 
can be obtained by interchanging $L$ and $R$ in the basis given above in 
model III. However, we do not present them here since corresponding  Wilson 
coefficients are negligible.
\section{The Initial values of the Wilson coefficients.}
The initial values of the Wilson coefficients for the relevant process 
in the SM are \cite{Grinstein2}
\begin{eqnarray}
C^{SM}_{1,3,\dots 6}(m_W)&=&0 \nonumber \, \, , \\
C^{SM}_2(m_W)&=&1 \nonumber \, \, , \\
C_7^{SM}(m_W)&=&\frac{3 x_t^3-2 x_t^2}{4(x_t-1)^4} \ln x_t+
\frac{-8 x_t^3-5 x_t^2+7 x_t}{24 (x_t-1)^3} \nonumber \, \, , \\
C_8^{SM}(m_W)&=&-\frac{3 x_t^2}{4(x_t-1)^4} \ln x_t+
\frac{-x_t^3+5 x_t^2+2 x_t}{8 (x_t-1)^3}\nonumber \, \, , \\ 
C_9^{SM}(m_W)&=&-\frac{1}{sin^2\theta_{W}} B(x_t) +
\frac{1-4 \sin^2 \theta_W}{\sin^2 \theta_W} C(x_t)-D(x_t)+
\frac{4}{9}\nonumber \, \, , \\
C_{10}^{SM}(m_W)&=&\frac{1}{\sin^2\theta_W}
(B(x_t)-C(x_t))\nonumber \,\, , \\
C_{Q_i}^{SM}(m_W) & = & 0~~~ i=1,..,10
\end{eqnarray}
and for the additional part due to charged Higgs bosons are 
\begin{eqnarray}
C^{H}_{1,\dots 6 }(m_W)&=&0 \nonumber \, , \\
C_7^{H}(m_W)&=& Y^2 \, F_{1}(y_t)\, + \, X Y \,  F_{2}(y_t) 
\nonumber  \, \, , \\
C_8^{H}(m_W)&=& Y^2 \,  G_{1}(y_t) \, + \, X Y \, G_{2}(y_t) 
\nonumber\, \, , \\
C_9^{H}(m_W)&=&  Y^2 \,  H_{1}(y_t) \nonumber  \, \, , \\
C_{10}^{H}(m_W)&=& Y^2 \,  L_{1}(y_t)  
\label{CH} \, \, , 
\end{eqnarray}
where 
\bea
X & = & \frac{1}{m_{b}}~~~\left(\bar{\xi}^{D}_{N,bb}+\bar{\xi}^{D}_{N,sb}
\frac{V_{ts}}{V_{tb}} \right) ~~,~~ \nnb \\
Y & = & \frac{1}{m_{t}}~~~\left(\bar{\xi}^{U}_{N,tt}+\bar{\xi}^{U}_{N,tc}
\frac{V^{*}_{cs}}{V^{*}_{ts}} \right) ~~.~~
\eea
The NHB effects bring new operators and the corresponding Wilson
coefficients  read as \cite{Ergur} 
\bea
%\!\!\!\!\!\!\!\!\!\!\!\!\!\!\!\!\!\!\!\!\!\!\!\!\!\!\!\!\!\!\!\!\!\!\!\!\!\!\!
%\!\!\!\!\!\!\!\!\!\!\!\!\!\!\!\!\!\!\!\!\!\!\!\!\!\!\!\!\!\!\!\!\!\!\!\!\!\!\!
C^{A^{0}}_{Q_{2}}((\bar{\xi}^{U}_{N,tt})^{3}) & = &
\frac{\bar{\xi}^{D}_{N,\tau \tau}(\bar{\xi}^{U}_{N,tt})^{3}
m_{b} y_t (\Theta_5 (y_t)z_A-\Theta_1 (z_{A},y_t))}{32 \pi^{2}m_{A^{0}}^{2} 
m_{t} \Theta_1 (z_{A},y_t) \Theta_5 (y_t)} , \nnb \\
C^{A^{0}}_{Q_{2}}((\bar{\xi}^{U}_{N,tt})^{2})& = & 
\frac{\bar{\xi}^{D}_{N,\tau\tau}(\bar{\xi}^{U}_{N,tt})^{2}
\bar{\xi}^{D}_{N,bb}}{32 \pi^{2}  m_{A^{0}}^{2}}\Big{(} 
\frac{1}{\Theta_1 (z_{A},y_t) \Theta_1 (z_{A},y_t) \Theta_5 (y_t)}\Big) \nnb
\\ & \cdot & (y_t (\Theta_1 (z_{A},y_t) - \Theta_5 (y_t) (xy+z_A))-
2 \Theta_1 (z_{A},y_t) \Theta_5 (y_t)   \ln [\frac{z_A \Theta_5
(y_t)}{\Theta_1(z_A,y_t)}]) \nnb
\eea
\bea
C^{A^{0}}_{Q_{2}}(\bar{\xi}^{U}_{N,tt}) &=& 
\frac{g^2\bar{\xi}^{D}_{N,\tau\tau}\bar{\xi}^{U}_{N,tt} m_b
x_t}{64 \pi^2 m_{A^{0}}^{2}  m_t } \Bigg{(}\frac{2}{\Theta_5 (x_t)}
- \frac{xy x_t+2 z_A}{\Theta_1 (z_{A},x_t)}-2
\ln [\frac{z_A \Theta_5(x_t)}{ \Theta_1 (z_{A},x_t)}]\nnb \\ &- & 
x y x_t y_t(\frac{(x-1) x_t 
(y_t/z_A-1)-(1+x)y_t)}{(\Theta_6 -(x-y)(x_t -y_t))(\Theta_3
(z_A)+(x-y)(x_t-y_t)z_A)} \nnb \\ & - &
\frac{x (y_t+x_t(1-y_t/z_A))-2 y_t }{\Theta_6 \Theta_3 (z_A)}) \Bigg{)} \nnb
\eea
\bea
\!\!\!C^{A^{0}}_{Q_{2}}(\bar{\xi}^{D}_{N,bb})& = &
\frac{g^2\bar{\xi}^{D}_{N,\tau\tau}\bar{\xi}^{D}_{N,bb}}{64 \pi^2 m^2_{A^{0}} }
\Big{(}1-
\frac{x^2_t y_t+2 y (x-1)x_t y_t-z_A (x^2_t+\Theta_6)}{ \Theta_3 (z_A)} \nnb
\\ & + &
\frac{x^2_t (1-y_t/z_A)}{\Theta_6}+2 \ln [\frac{z_A \Theta_6}
{ \Theta_2 (z_{A},x)}] \Big{)}\nnb 
\eea
\bea
\lefteqn{\!\!\!\!\!\!\!\!\!\!\!\!\!\!\!\!\!\!\!\!\!C^{H^{0}}_{Q_{1}}
((\bar{\xi}^{U}_{N,tt})^{2}) = 
\frac{g^2 (\bar{\xi}^{U}_{N,tt})^2 m_b m_{\tau}
}{64 \pi^2 m^2_{H^{0}} m^2_t } \Bigg{(}
\frac{x_t (1-2 y) y_t}{\Theta_5 (y_t)}+\frac{(-1+2 \cos^2 \theta_W) (-1+x+y) 
y_t} {\cos^2 \theta_W \Theta_4 (y_t)} } \nnb
\\ & &
+\frac{z_H (\Theta_1 (z_H,y_t) x y_t + 
\cos^2 \theta_W \,(-2 x^2 (-1+x_t) y y^2_t+x x_t y y^2_t-\Theta_8 z_H))}
{\cos^2 \theta_W \Theta_1 (z_H,y_t) \Theta_7 }\Bigg{)} ,  \nnb
\eea
\bea
C^{H^{0}}_{Q_{1}}
 (\bar{\xi}^{U}_{N,tt}) 
 & = & \frac{g^2 \bar{\xi}^{U}_{N,tt} \bar{\xi}^{D}_{N,bb} 
m_{\tau}}{64 \pi^2 m^2_{H^{0}} m_t } \Bigg{(}
\frac{(-1+2 \cos^2 \theta_W)\, y_t}{\cos^2 \theta_W \, \Theta_4 (y_t)}-
\frac{x_t y_t}{\Theta_5 (y_t)}+\frac {x_t y_t(x y-z_H)}
{\Theta_1 (z_H,y_t)}  \nnb \\
& + & \frac{(-1+2 \cos^2 \theta_W) 
y_t z_H}{\cos^2\theta_W \Theta_7}-2 x_t\, \ln \Bigg{[}
\frac{\Theta_5 (y_t) z_H} {\Theta_1 (z_H,y_t)} \Bigg{]} \Bigg{)}  ,
\label{NHB}
\eea
\bea
\lefteqn{ C^{H^0}_{Q_{1}}(g^4) =-\frac{g^4 m_b m_{\tau} x_t}
{128 \pi^2 m^2_{H^{0}} m^2_t} 
\Bigg{(} -1+\frac{(-1+2x) x_t}{\Theta_5 (x_t) + y (1-x_t)}+
\frac{2 x_t (-1+ (2+x_t) y)}{\Theta_5 (x_t)} } \nnb \\
& & 
-\frac{4 \cos^2 \theta_W (-1+x+y)+ x_t(x+y)} {\cos^2 \theta_W 
\Theta_4 (x_t)} +\frac{x_t (x (x_t (y-2 z_H)-4 z_H)+2 z_H)} {\Theta_1 
(z_H,x_t)} \nnb \\ 
& &
+\frac{y_t ( (-1+x) x_t z_H+\cos^2 \theta_W ( (3 x-y) z_H+x_t 
(2 y (x-1)- z_H (2-3 x -y))))}{\cos^2 \theta_W (\Theta_3 (z_H)+x 
(x_t-y_t) z_H)} \nnb
\\ & & 
+ 2\, ( x_t \ln \Bigg{[} \frac{\Theta_5 (x_t) z_H}{\Theta_1 (z_H,x_t)} 
\Bigg{]}+ \ln \Bigg{[} \frac{x(y_t-x_t) z_H-\Theta_3 (z_H)} {(\Theta_5 (x_t)+ 
y (1-x_t) y_t z_H} \Bigg{]} )\Bigg{)}  ,\nnb  
\eea
\bea
C^{h_0}_{Q_1}((\bar{\xi}^U_{N,tt})^3) &=&
-\frac{\bar{\xi}^D_{N,\tau\tau} (\bar{\xi}^U_{N,tt})^3 m_b y_t}
{32 \pi^2 m_{h^0}^2 m_t \Theta_1 (z_h,y_t) \Theta_5 (y_t)}
 \Big{(} \Theta_1 (z_h,y_t) (2 y-1) + \Theta_5 (y_t) (2 x-1) z_h \Big{)} \nnb 
\eea
\bea
C^{h_0}_{Q_1}((\bar{\xi}^U_{N,tt})^2) & = &
\frac{\bar{\xi}^D_{N,\tau\tau} (\bar{\xi}^U_{N,tt})^2 }
{32 \pi^2  m_{h^0}^2  } \Bigg{(}
\frac{ (\Theta_5 (y_t) z_h (y_t-1)(x+y-1)-\Theta_1 (z_h,y_t)( \Theta_5(y_t)+y_t )
}{\Theta_1 (z_h)\Theta_5(y_t)} \nnb \\ &-& 2 \ln \Bigg{[} \frac{z_h \Theta_5
(y_t)}{\Theta_1 (z_h)} \Bigg{]} \Bigg{)}\nnb 
\eea
\bea
C^{h^0}_{Q_{1}}(\bar{\xi}^{U}_{N,tt}) & = & -\frac{g^2
\bar{\xi}^{D}_{N,\tau\tau}\bar{\xi}^{U}_{N,tt} m_b x_t}{64 \pi^2 m^2_{h^{0}} 
m_t} \Bigg{(}\frac{2 (-1+(2+x_t) y)}{\Theta_5 (x_t)}-\frac{x_t
(x-1)(y_t-z_h)}{\Theta'_2 (z_h)}+2 \ln \Bigg{[}\frac{z_h \Theta_5
(x_t)}{\Theta_1 (z_h,x_t)} \Bigg{]} \nnb \\ & + & \frac{x (x_t (y-2 z_h)-
4 z_h)+2 z_h}{\Theta_1 (z_h,x_t)}  -  \frac{(1+x) y_t z_h}{x y x_t y_t+z_h
((x-y)(x_t-y_t)- \Theta_6)} \nnb \\ 
& + & 
\frac{\Theta_9 + y_t z_h ( (x-y)(x_t-y_t)-\Theta_6 )(2
x-1)}{z_h \Theta_6 (\Theta_6 -(x-y)(x_t-y_t))}+\frac{x (y_t z_h + x_t
(z_h-y_t))-2 y_t z_h}{\Theta_2 (z_h)} \Bigg{)}, \nnb
\eea
\bea
C^{h^0}_{Q_{1}}(\bar{\xi}^{D}_{N,bb}) &  = & -\frac{g^2
\bar{\xi}^{D}_{N,\tau\tau}\bar{\xi}^{D}_{N,bb}}{64 \pi^2 m^2_{h^0} }
\Bigg{(}\frac{y x_t y_t (x x^2_t(y_t-z_h)+\Theta_6 z_h
(x-2))}{z_h\Theta_2 (z_h)\Theta_6 }+2 \ln \Bigg{[}\frac{\Theta_6}{x_t y_t} \Bigg{]}
\nnb \\ &+&2 \ln \Bigg{[}\frac{x_t y_t z_h}{\Theta_2 (z_h)} \Bigg{]}
\Bigg{)} \nnb 
\eea
where 
\bea
\Theta_1 (\omega , \lambda ) & = & -(-1+y-y \lambda ) \omega -x (y \lambda
+\omega - \omega \lambda ) \nnb \\
\Theta_2 (\omega ) & = &  (x_t +y (1-x_t)) y_t \omega - x x_t (y
y_t+(y_t-1) \omega)   \nnb \\
\Theta^{\prime}_2  (\omega ) & = & \Theta_2 (\omega , x_t \leftrightarrow y_t)    \nnb \\
\Theta_3 (\omega) & = & (x_t (-1+y)-y ) y_t \omega +
x x_t (y y_t+\omega(-1+y_t)) \nnb \\
\Theta_4 (\omega) & = & 1-x +x  \omega  \nnb \\
\Theta_5 (\lambda) & = & x + \lambda (1-x) \nnb \\
\Theta_6  & = & (x_t +y  (1-x_t))y_t +x x_t  (1-y_t) \nnb \\
\Theta_7  & = & (y (y_t -1)-y_t) z_H+x (y y_t + (y_t-1) z_H ) \\ 
\Theta_8  & = & y_t (2 x^2 (1+x_t) (y_t-1) +x_t (y(1-y_t)+y_t)+x
(2(1-y+y_t) \nnb \\ & + & x_t (1-2 y (1-y_t)-3 y_t))) \nnb \\
\Theta_9  & = & -x^2_t (-1+x+y)(-y_t+x (2 y_t-1)) (y_t-z_h)-x_t y_t z_h
(x(1+2 x)-2 y) \nnb \\ & + & y^2_t (x_t (x^2 -y (1-x))+(1+x) (x-y) z_h) 
\nnb
\eea
and
\begin{eqnarray}
& & x_t=\frac{m_t^2}{m_W^2}~~~,~~~y_t=
\frac{m_t^2}{m_{H^{\pm}}}~~~,~~~z_H=\frac{m_t^2}{m^2_{H^0}}~~~,~~~
z_h=\frac{m_t^2}{m^2_{h^0}}~~~,~~~ z_A=\frac{m_t^2}{m^2_{A^0}}~~~.~~~ \nnb
\end{eqnarray}
The explicit forms of the functions $F_{1(2)}(y_t)$, $G_{1(2)}(y_t)$, 
$H_{1}(y_t)$ and $L_{1}(y_t)$ in Eq.(\ref{CH}) are given as
\begin{eqnarray}
F_{1}(y_t)&=& \frac{y_t(7-5 y_t-8 y_t^2)}{72 (y_t-1)^3}+
\frac{y_t^2 (3 y_t-2)}{12(y_t-1)^4} \,\ln y_t \nonumber  \,\, , 
\\ 
F_{2}(y_t)&=& \frac{y_t(5 y_t-3)}{12 (y_t-1)^2}+
\frac{y_t(-3 y_t+2)}{6(y_t-1)^3}\, \ln y_t 
\nonumber  \,\, ,
\\ 
G_{1}(y_t)&=& \frac{y_t(-y_t^2+5 y_t+2)}{24 (y_t-1)^3}+
\frac{-y_t^2} {4(y_t-1)^4} \, \ln y_t
\nonumber  \,\, ,
\\ 
G_{2}(y_t)&=& \frac{y_t(y_t-3)}{4 (y_t-1)^2}+\frac{y_t} {2(y_t-1)^3} \, 
\ln y_t  \nonumber\,\, ,
\\
H_{1}(y_t)&=& \frac{1-4 sin^2\theta_W}{sin^2\theta_W}\,\, \frac{xy_t}{8}\,
\left[ \frac{1}{y_t-1}-\frac{1}{(y_t-1)^2} \ln y_t \right]\nonumber \\
&-&
y_t \left[\frac{47 y_t^2-79 y_t+38}{108 (y_t-1)^3}-
\frac{3 y_t^3-6 y_t+4}{18(y_t-1)^4} \ln y_t \right] 
\nonumber  \,\, , 
\\ 
L_{1}(y_t)&=& \frac{1}{sin^2\theta_W} \,\,\frac{x y_t}{8}\, 
\left[-\frac{1}{y_t-1}+ \frac{1}{(y_t-1)^2} \ln y_t \right]
\nonumber  \,\, .
\\ 
\label{F1G1}
\end{eqnarray}
Finally, the initial values of the coefficients in the model III are
\begin {eqnarray}   
C_i^{2HDM}(m_{W})&=&C_i^{SM}(m_{W})+C_i^{H}(m_{W}) , \nnb \\
C_{Q_{1}}^{2HDM}(m_{W})&=& \int^{1}_{0}dx \int^{1-x}_{0} dy \,
(C^{H^{0}}_{Q_{1}}((\bar{\xi}^{U}_{N,tt})^{2})+
 C^{H^{0}}_{Q_{1}}(\bar{\xi}^{U}_{N,tt})+
 C^{H^{0}}_{Q_{1}}(g^{4})+C^{h^{0}}_{Q_{1}}((\bar{\xi}^{U}_{N,tt})^{3}) \nnb
\\ & + &
 C^{h^{0}}_{Q_{1}}((\bar{\xi}^{U}_{N,tt})^{2})+
 C^{h^{0}}_{Q_{1}}(\bar{\xi}^{U}_{N,tt})+
 C^{h^{0}}_{Q_{1}}(\bar{\xi}^{D}_{N,bb})) , \nnb  \\
 C_{Q_{2}}^{2HDM}(m_{W})&=& \int^{1}_{0}dx \int^{1-x}_{0} dy\,
(C^{A^{0}}_{Q_{2}}((\bar{\xi}^{U}_{N,tt})^{3})+
C^{A^{0}}_{Q_{2}}((\bar{\xi}^{U}_{N,tt})^{2})+
 C^{A^{0}}_{Q_{2}}(\bar{\xi}^{U}_{N,tt})+
 C^{A^{0}}_{Q_{2}}(\bar{\xi}^{D}_{N,bb}))\nnb \\
C_{Q_{3}}^{2HDM}(m_W) & = & \frac{m_b}{m_{\tau} \sin^2 \theta_W} 
 (C_{Q_{1}}^{2HDM}(m_W)+C_{Q_{2}}^{2HDM}(m_W)) \nnb \\
C_{Q_{4}}^{2HDM}(m_W) & = & \frac{m_b}{m_{\tau} \sin^2 \theta_W} 
 (C_{Q_{1}}^{2HDM}(m_W)-C_{Q_{2}}^{2HDM}(m_W)) \nnb \\
C_{Q_{i}}^{2HDM} (m_W) & = & 0\,\, , \,\, i=5,..., 10.
\label{CiW}
\end{eqnarray}
Here, we present $C_{Q_{1}}$ and $C_{Q_{2}}$ in terms of the Feynman
parameters $x$ and $y$ since the integrated results are extremely large.
Using these initial values, we can calculate the coefficients 
$C_{i}^{2HDM}(\mu)$ and $C^{2HDM}_{Q_i}(\mu)$ 
at any lower scale in the effective theory 
with five quarks, namely $u,c,d,s,b$ similar to the SM case 
\cite{Alil2,Misiak,Chao,Buras}. 

The Wilson  coefficients playing  the essential role 
in this process are $C_{7}^{2HDM}(\mu)$, $C_{9}^{2HDM}(\mu)$,
$C_{10}^{2HDM}(\mu)$, 
$C^{2HDM}_{Q_1}(\mu )$ and $C^{2HDM}_{Q_2}(\mu )$. For completeness,
in the following we give their explicit expressions,
\begin{eqnarray}
C_{7}^{eff}(\mu)&=&C_{7}^{2HDM}(\mu)+ Q_d \, 
(C_{5}^{2HDM}(\mu) + N_c \, C_{6}^{2HDM}(\mu))\nonumber \, \, ,
\label{C7eff}
\end{eqnarray}
where the LO  QCD corrected Wilson coefficient 
$C_{7}^{LO, 2HDM}(\mu)$ is given by
\begin{eqnarray} 
C_{7}^{LO, 2HDM}(\mu)&=& \eta^{16/23} C_{7}^{2HDM}(m_{W})+(8/3) 
(\eta^{14/23}-\eta^{16/23}) C_{8}^{2HDM}(m_{W})\nonumber \,\, \\
&+& C_{2}^{2HDM}(m_{W}) \sum_{i=1}^{8} h_{i} \eta^{a_{i}} \,\, , 
\label{LOwils}
\end{eqnarray}
and $\eta =\alpha_{s}(m_{W})/\alpha_{s}(\mu)$, $h_{i}$ and $a_{i}$ are 
the numbers which appear during the evaluation \cite{Buras}. 

$C_9^{eff}(\mu)$ contains a perturbative part and a part coming from LD
effects due to conversion of the real $\bar{c}c$ into lepton pair $\ell^+
\ell^-$:
\begin{eqnarray}
C_9^{eff}(\mu)=C_9^{pert}(\mu)+ Y_{reson}(s)\,\, ,
\label{C9efftot}
\end{eqnarray}
where
\begin{eqnarray} 
C_9^{pert}(\mu)&=& C_9^{2HDM}(\mu) \nonumber 
\\ &+& h(z,  s) \left( 3 C_1(\mu) + C_2(\mu) + 3 C_3(\mu) + 
C_4(\mu) + 3 C_5(\mu) + C_6(\mu) \right) \nonumber \\
&- & \frac{1}{2} h(1, s) \left( 4 C_3(\mu) + 4 C_4(\mu) + 3
C_5(\mu) + C_6(\mu) \right) \\
&- &  \frac{1}{2} h(0,  s) \left( C_3(\mu) + 3 C_4(\mu) \right) +
\frac{2}{9} \left( 3 C_3(\mu) + C_4(\mu) + 3 C_5(\mu) + C_6(\mu)
\right) \nonumber \,\, ,
\label{C9eff2}
\end{eqnarray}
and
\begin{eqnarray}
Y_{reson}(s)&=&-\frac{3}{\alpha^2_{em}}\kappa \sum_{V_i=\psi_i}
\frac{\pi \Gamma(V_i\rightarrow \ell^+ \ell^-)m_{V_i}}{q^2-m_{V_i}+i m_{V_i}
\Gamma_{V_i}} \nonumber \\
& & \left( 3 C_1(\mu) + C_2(\mu) + 3 C_3(\mu) + 
C_4(\mu) + 3 C_5(\mu) + C_6(\mu) \right).
\label{Yres}
\end{eqnarray}
In eq.(\ref{C9efftot}), the functions $h(u, s)$ are given by
\begin{eqnarray}
h(u, s) &=& -\frac{8}{9}\ln\frac{m_b}{\mu} - \frac{8}{9}\ln u +
\frac{8}{27} + \frac{4}{9} x \\
& & - \frac{2}{9} (2+x) |1-x|^{1/2} \left\{\begin{array}{ll}
\left( \ln\left| \frac{\sqrt{1-x} + 1}{\sqrt{1-x} - 1}\right| - 
i\pi \right), &\mbox{for } x \equiv \frac{4u^2}{ s} < 1 \nonumber \\
2 \arctan \frac{1}{\sqrt{x-1}}, & \mbox{for } x \equiv \frac
{4u^2}{ s} > 1,
\end{array}
\right. \\
h(0,s) &=& \frac{8}{27} -\frac{8}{9} \ln\frac{m_b}{\mu} - 
\frac{4}{9} \ln s + \frac{4}{9} i\pi \,\, , 
\label{hfunc}
\end{eqnarray}
with $u=\frac{m_c}{m_b}$.
The phenomenological parameter $\kappa$ in eq. (\ref{Yres}) is taken as 
$2.3$. In Eqs. (37) and (\ref{Yres}), the contributions of 
the coefficients $C_1(\mu)$, ...., $C_6(\mu)$ are due to the operator mixing.

Finally, the Wilson coefficients $C_{Q_1}(\mu)$ and $C_{Q_2}(\mu )$  
are given by \cite{Dai}
\beq
C_{Q_i}(\mu )=\eta^{-12/23}\,C_{Q_i}(m_W)~,~i=1,2~. 
\eeq
\section{Some functions appearing in the expressions}
We parametrize the functions $A_1$, $A_2$, $B_1$, $B_2$, $F_1$ and  $F_2$ in
Eqs. (\ref{Msd}) and (\ref{Mib}) as 
\begin{eqnarray}
A_1&=&A^{(1)}_1 + i\, A^{(2)}_1 + \bar{\xi}^D_{N,bb}\, A^{(3)}_1 
\nnb \,\, \\
A_2&=&A^{(1)}_2 + i\, A^{(2)}_2 + \bar{\xi}^D_{N,bb}\, A^{(3)}_2  
\nnb \,\, \\
B_1 & = & C_{10} \, g \, , \nnb \\
B_2 & = & C_{10} \, f \, , \nnb \\
F &=& 2 m_{\tau} C_{10} +\frac{m^2_B}{m^2_b} \, \left(F^{(1)} + \bar{\xi}^D_{N,bb}\, 
F^{(2)} \right) \nonumber \,\, , \\
F_1 &=& \frac{m^2_B}{m^2_b} \, \left(F^{(1)}_1 + \bar{\xi}^D_{N,bb}\,
F^{(2)}_1 \right) \, ,\label{ABFF1}
\end{eqnarray}
with 
\begin{eqnarray}
A^{(1)}_1&=& g \, \mbox{\rm Re}  (C_9^{eff}) - \frac{2 m_b }{q^2}\,g_1 \, 
C_7^{eff} \Big{|}_{\bar{\xi}^D_{N,bb}\rightarrow 0} \nonumber \,\, , \\
A^{(2)}_1 &=& g\, \mbox{\rm Im}(C_9^{eff})  \nonumber \,\, , \\
A^{(3)}_1 &=& - \frac{2 m_b}{q^2} \,g_1 \,\frac{1}{m_b\,m_t} \bar{\xi}^U_{N,tt}
(\eta^{\frac{16}{23}} K_2 (y_t)+\frac{8}{3}
(\eta^{\frac{14}{23}}-\eta^{\frac{16}{23}}) G_2 (y_t))
\nonumber \,\, , \\
A^{(1)}_2 &=& A^{(1)}_1 (g\rar f;g_1 \rar f_1) \nnb \, , \\
A^{(2)}_2 &=& A^{(2)}_1 (g\rar f) \nnb \, , \\
A^{(3)}_2 &=& A^{(3)}_1 (g_1 \rar f_1) \nnb \, , \\
F_1^{(1)}&=&\eta^{-12/23}\,\int^{1}_{0}dx \int^{1-x}_{0} dy \,
[C^{H_0}_{Q_{1}}((\bar{\xi}^{U}_{N,tt})^{2})+
C^{H_0}_{Q_{1}}(\bar{\xi}^{U}_{N,tt})
\nonumber \\ &+&
C^{H_0}_{Q_{1}}(g^{4})+C^{h_0}_{Q_{1}}((\bar{\xi}^{U}_{N,tt})^{3}) 
+ C^{h_0}_{Q_{1}}((\bar{\xi}^{U}_{N,tt})^{2})+
C^{h_0}_{Q_{1}}(\bar{\xi}^{U}_{N,tt}) ] \nonumber \,\, , \\
F_1^{(2)}&=&  \frac{\eta^{-12/23}}{\bar{\xi}^{D}_{N,bb}}\, \int^{1}_{0}dx \int^{1-x}_{0}
 dy \,\,C^{h_0}_{Q_{1}}(\bar{\xi}^{D}_{N,bb}) \nonumber \,\, , \\
F^{(1)}&=&\eta^{-12/23}\, \int^{1}_{0}dx \int^{1-x}_{0} dy\,
\left[C^{A_0}_{Q_{2}}((\bar{\xi}^{U}_{N,tt})^{3})+
C^{A_0}_{Q_{2}}((\bar{\xi}^{U}_{N,tt})^{2})+
C^{A_0}_{Q_{2}}(\bar{\xi}^{U}_{N,tt}) \right] \nonumber \,\, , \\
F^{(2)}&=&\frac{\eta^{-12/23}}{\bar{\xi}^{D}_{N,bb}}\, \int^{1}_{0}dx \int^{1-x}_{0} dy\,
C^{A_0}_{Q_{2}}(\bar{\xi}^{D}_{N,bb})\,\, . \label{AF1F2b}
\end{eqnarray}
\end{appendix}
\newpage
\end{document}